\documentclass[a4paper,11pt]{article}
\usepackage{amsthm}
\usepackage{amsmath}
\usepackage{amssymb}
\usepackage{color}
\usepackage{cite}
\usepackage{graphicx}
\usepackage{xcolor}
\usepackage{subfigure}
\usepackage[linktocpage]{hyperref}
\usepackage{setspace}
\usepackage{comment}

\usepackage{titletoc}
\dottedcontents{section}[1.2em]{}{1em}{0.5pc}
\dottedcontents{subsection}[4em]{}{2em}{1pc}

\usepackage{etoolbox}
\makeatletter
\pretocmd{\subsection}{\addtocontents{toc}{\protect\addvspace{-4\p@}}}{}{}
\pretocmd{\section}{\addtocontents{toc}{\protect\addvspace{-4\p@}}}{}{}
\makeatother
\newcommand{\mrd}{\mathrm{d}}
\renewcommand{\H}{{_\mathrm{H}}}

\definecolor{darkred}{rgb}{0.9,0.05,0.05}
\definecolor{darkblue}{rgb}{0.05,0.05,0.6}
\definecolor{darkgreen}{rgb}{0.05,0.6,0.05}
\definecolor{brightgreen}{rgb}{0.1,0.9,0.1}

\renewcommand*{\eqref}[1]{%
  \begingroup
    \hypersetup{
      linkcolor=darkblue,
      linkbordercolor=darkblue,
    }%
    \textcolor{darkblue}{(\ref{#1})}%
  \endgroup 
}

\hypersetup{linkcolor=red,citecolor=brightgreen,urlcolor=black,colorlinks=true}

\begin{document}
\let\endtitlepage\relax

\begin{titlepage}
\begin{center}
\renewcommand{\baselinestretch}{1.5}  
\setstretch{1.5}

\vspace*{-0.5cm}

{\fontsize{19pt}{22pt}\bf{First law of black hole thermodynamics and Smarr formula with a cosmological constant}}
 
\vspace{9mm}
\renewcommand{\baselinestretch}{1}  
\setstretch{1}

\centerline{\large{Kamal  Hajian$^{\dagger\ddagger}$}\footnote{khajian@metu.edu.tr}, \large{Hikmet Özşahin$^\dagger$}\footnote{hikmet.ozsahin@metu.edu.tr}, \large{Bayram Tekin$^\dagger$}\footnote{btekin@metu.edu.tr}}

\vspace{5mm}
\normalsize
\textit{$^\dagger$Department of Physics, Middle East Technical University, 06800, Ankara, Turkey}\\
\textit{$^\ddagger$Hanse-Wissenschaftskolleg Institute for Advanced Study, Lehmkuhlenbusch 4, 27733 Delmenhorst, Germany}
\vspace{5mm}

\begin{abstract}
The first law of black hole thermodynamics in the presence of a cosmological constant $\Lambda$ can be generalized  by introducing a term containing the variation $\delta\Lambda$. Similar to other terms in the first law, which are variations of some conserved charges like mass, entropy, angular momentum, electric charge etc., it has been shown [Classical Quantum Gravity 35, 125012 (2018)] that the new term has the same structure: $\Lambda$ is a conserved charge associated with a gauge symmetry; and its role in the first law is quite similar to an ``electric charge" rather than to the pressure. Besides, its conjugate chemical potential resembles an ``electric potential" on the horizon, in contrast with the volume enclosed by horizon. In this work, first we propose and prove the generalized Smarr relation in  this new paradigm. Then, we reproduce systematically the ``effective volume" of a black hole which has been introduced before in the literature as the conjugate of pressure. Our construction removes the ambiguity in the definition of volume. Finally, we apply and investigate this formulation of ``$\Lambda$ as a charge" on a number of solutions to different models of gravity for different spacetime dimensions. Especially, we investigate the applicability and validity of the analysis for black branes, whose enclosed volume is not well defined in principle.      

\end{abstract}
\end{center}
\vspace*{0cm}


\setcounter{tocdepth}{2}

 \tableofcontents
\end{titlepage}
\vspace*{0cm}

\renewcommand{\baselinestretch}{1.05}  
\setstretch{1}
\section{Introduction}
Einstein introduced the cosmological constant $\Lambda$ in order to explain, within general relativity, the ``apparent" staticity nature of the Universe  \cite{Einstein:1917}. But, after the discovery of the  expansion of the Universe in the late 1920s, the idea of a static universe was essentially put to rest  together with the cosmological constant by the majority of the researchers in the field. However, in a rather ironic piece of scientific history, $\Lambda$ has taken center stage in cosmology since the discovery of accelerating expansion of the Universe \cite{Perlmutter:1998np,Riess:1998cb} and the AdS/CFT correspondence \cite{Maldacena:1997re,Brown:1986nw}. In the Einstein's method of introducing the cosmological constant, $\Lambda$ is considered as a constant parameter in the Lagrangian, {\it i.e.} as a part of the definition of the theory. Mathematically, denoting the Lagrangian by $\mathcal{L}$,  $\Lambda$ is incorporated in the Lagrangian via the shift $\mathcal{L}\to \mathcal{L}-\frac{\Lambda}{8\pi G}$ where $G$ is the Newton constant.  {Alternatively, in a less-known route, one can introduce a new gauge field in the Lagrangian \cite{Aurilia:1980xj, Duff:1980qv}  which makes $\Lambda$ to be a free parameter in the solution. This approach was introduced in the early 1980s, and was studied in more detail in a series of papers by Henneaux and  Teitelboim \cite{Henneaux:1984ji,Henneaux:1985tv,Teitelboim:1985dp,Henneaux:1989zc}, who studied Hamiltonian dynamics of this new gauge field and identified its canonical variables (canonical field and its momentum conjugate), and constants of integration. Continuing this research line, it is shown that not only $\Lambda$ is a constant of integration in the solution, but also (its square-root) is a conserved charge (denoted by $C$) associated with the global part of the gauge symmetry of this gauge field \cite{1}. In addition, its conjugate chemical potential associated with a black hole horizon (denoted by $\Theta_\H$) was introduced for the first time. This formulation brings a new perspective to  $\Lambda$: It becomes a conserved charge as a property of the solution and can naturally contribute to the first law of black hole thermodynamics just like other conserved charges. This can also be considered as a continuation of the seminal work by Wald \cite{Wald:1993nt,Iyer:1994ys} who recognized the the entropy in the black hole thermodynamics is a conserved charge.}  Consequently, this approach resolves some conceptual, physical, and mathematical issues in regard to the generalization of the first law with variation of the cosmological constant and the issues with the Smarr formula. We will come back and summarize these issues later in this section.

For the sake of completeness, in what follows, we briefly review  the ``$\Lambda$ as a conserved charge" approach \cite{1}. We shall use the following conventions: 
$[\mu_1\mu_2\dots\mu_p]$ will be used to denote antisymmetrization over the set of indices within the bracket normalized by the factor $\frac{1}{p!}$. The exterior derivative of a $p$-form $\boldsymbol{a}=\frac{1}{p!}a_{\rho_1\dots \rho_p}dx^{\rho_1}\wedge\dots\wedge dx^{\rho_p}$ is defined as
\begin{equation}
\mrd\boldsymbol{a} \equiv (p+1)\,\partial_{[\mu_1}a_{\mu_2\dots\mu_{p+1}]}\,\mrd x^{\mu_1}\wedge\dots\wedge \mrd x^{\mu_{p+1}}.\nonumber
\end{equation}

{Considering a gravitational theory described by a Lagrangian $\mathcal{L}$ without a cosmological constant in $D$-dimensional spacetime, the action and gravitational equation of motion can be represented as
\begin{equation}
I=\int \mrd^D x \sqrt{-g}\mathcal{L}, \qquad \qquad E_{\mu\nu}\equiv \frac{\delta (\sqrt{-g}\mathcal{L})}{\delta g^{\mu\nu}}=0,    
\end{equation}
in which $\delta g^{\mu\nu}$ is variations of the inverse metric. In order to introduce a cosmological constant, one can add a gauge field Lagrangian (a term similar to the electromagnetic Lagrangian) to the gravity sector as
\begin{equation}\label{L F2}
\mathcal{L}\to \mathcal{L}\mp\frac{1}{8\pi G} F^2 \quad \Rightarrow \quad  I=\int \mrd^D x \sqrt{-g}\Big(\mathcal{L}  \mp\frac{1}{8\pi G} F^2 \Big),
\end{equation}
where $F^2 \equiv \frac{1}{D!}F_{\mu_1\dots\mu_D}F^{\mu_1\dots\mu_D}$. $F$ is a top-form (i.e., having $D$ antisymmetric indices) and is the field strength of a gauge field $F=\mrd A$, i.e.,
\begin{equation}
\frac{1}{D!}F_{\mu_1\dots\mu_D}=\partial_{[\mu_1} A_{\mu_2 \dots \mu_D]}.  
\end{equation}
 We note that the new term in the Lagrangian \eqref{L F2} is quite similar to the Maxwell Lagrangian, and the only difference is that $A$ and $F$ have $D-1$ and $D$ indices (instead of 1 and 2 indices) respectively. In general, the top-form  $F$ can be an arbitrary scalar function times the volume form, i.e. $F_{\mu_1\dots\mu_D}=\phi(x^\mu)\sqrt{-g}\epsilon_{\mu_1\dots\mu_D}$, with the convention $\epsilon_{01\dots D-1}=+1$ for the Levi-Civita tensor density. In another words, the most generic $F$ is Hodge dual to a scalar field $\phi$.  With variation of the action \eqref{L F2} with respect to $g_{\mu\nu}$ and $F$,one finds the following two field equations:
\begin{align}\label{EH equation}
&E_{\mu\nu}= \frac{\pm 1}{8\pi G(D-1)!} \Big(F_{\mu\rho_2\dots\rho_D}F_{\nu}{}^{\rho_2\dots\rho_D}-\frac{(D-1)!}{2} F^2 g_{\mu\nu}\Big),\\
&\nabla_\mu F^{\mu\mu_2\dots\mu_{D}}=0. \label{F eq}
\end{align}
The latter equation is easy to solve, and the result is
\begin{equation}\label{F on-shell}
F_{\mu_1\dots\mu_D}=c\sqrt{-g}\,\epsilon_{\mu_1\dots\mu_D}    
\end{equation}
for a constant $c$. We assume $0\leq c$ for later convenience; and $c$ should not be confused with the speed of light, which is set to 1. It is easy to see why \eqref{F on-shell} is the generic solution for the equation of motion \eqref{F eq}, because in terms of the Hodge dual field $\phi(x^\mu)$, \eqref{F eq} is simply $\mrd \phi(x^\mu)=0$, which admits $\phi(x^\mu)=c=\text{constant}$ as its most generic solution. 

The solution \eqref{F on-shell} can be put in the field equation \eqref{EH equation} in order to reproduce the standard field equation with a cosmological constant:
\begin{equation}\label{EoM Lambda}
E_{\mu\nu}+\frac{1}{16\pi G}\Lambda g_{\mu\nu}=0, \qquad\qquad  \Lambda =\pm c^2.
\end{equation}
To derive this equation, the identities $
\epsilon_{\mu_1\dots\mu_D}\epsilon^{\mu_1\dots\mu_D}=-D!$ and $\epsilon_{\mu\rho_2\dots\rho_D}\epsilon_{\nu}{}^{\rho_2\dots\rho_D}=-(D-1)!g_{\mu\nu}$
have been used, in which $\epsilon^{01\dots D-1}=-1$. This procedure of introducing $\Lambda$ as a parameter of the solution  (instead of a constant in the Lagrangian) can be applied in any gravity theory; {i.e.,} it is independent of the $\mathcal{L}$ in the analysis above.

In a $U(1)$ gauge theory with the gauge symmetry $A_\mu \to A_\mu+\partial_\mu \lambda(x^\mu)$, the conserved charge (such as the electric charge) is associated with the global part of the symmetry $\partial_\mu \lambda=0$.  Similarly, the Lagrangian \eqref{L F2} has a gauge symmetry 
\begin{equation}
A_{\mu_1 \dots \mu_{D-1}} \to A_{\mu_1 \dots \mu_{D-1}}+\partial_{[\mu_1}\lambda_{\mu_2 \dots \mu_{D-1}]}.
\end{equation}
It was shown in Ref. \cite{1} that  the conserved charge of the global part of this symmetry $\partial_{[\mu_1}\lambda_{\mu_2 \dots \mu_{D-1}]}=0$, which we denote as $C$, is equal to
\begin{align}\label{C}
C= \pm \frac{\sqrt{|\Lambda|}}{4\pi G}.   
\end{align}
The signs correspond to those in the Lagrangian \eqref{L F2} and are associated with de Sitter (upper sign which here is plus) and anti-de Sitter (lower sign which here is minus) sectors. These $\pm$ signs (upper and lower signs) and their correspondence with the dS and AdS sectors will continue to be valid in the rest of this work.  We shall call the conserved charge as \emph{cosmological charge} in order to distinguish $C$ from $\Lambda$ (which is called cosmological constant).
Moreover, we shall call the \emph{cosmological gauge field} and the \emph{cosmological field strength} for $A_{\mu_1 \dots \mu_{D-1}}$ and $F_{\mu_1 \dots \mu_{D}}$ respectively. 

Identification of $C$ as the cosmological charge turns out to be very useful in the black hole thermodynamics. The first law of thermodynamics for an electrically charged black hole in asymptotic flat spacetimes reads as 
$
\delta M=T_{_\text{H}}\delta S+\Omega_{_\text{H}}\delta J+\Phi_{_\text{H}}\delta Q  
$ \cite{Bardeen:1973gs}, where $(M,S,J,Q)$ are mass or energy, entropy, angular momentum, and electric charge of the black hole, respectively. All of these quantities, whose variations appear in the first law, are conserved charges associated with a symmetry. In addition, these quantities are all extensive thermodynamic quantities. On the other hand, $(T_{_\text{H}}, \Omega_{_\text{H}}, \Phi_{_\text{H}})$ are the temperature, angular velocity, and the electric potential of the black hole, respectively, all of which can be calculated using the metric on the black hole horizon, hence the subscript H. These quantities are all intensive quantities. Let us note that the electric potential is defined with $\Phi_{_\text{H}}\equiv \langle \xi_{_\text{H}},A \rangle =\xi_{_\text{H}}\cdot A  $ calculated on the horizon,  in which $\xi_{_\text{H}}$ is a horizon-generating null Killing vector field and $A$ is the electromagnetic gauge field $A_\mu$.  

According to the analysis above, $C$ is a new conserved charge for black hole solutions in asymptotically  (A)dS spacetimes,  which naturally should appear in the first law in equal footing with the other charges. This generalization has been elaborated in Ref. \cite{1}, and the modified first law reads
\begin{equation}\label{first law}
\delta M=T_{_\text{H}}\delta S+\Omega_{_\text{H}}\delta J+\Phi_{_\text{H}}\delta Q +\Theta_{_\text{H}}\delta C, \end{equation}
with 
\begin{equation}\label{Theta H}
\Theta_{_\text{H}}\equiv \oint_{\text{H}} \xi_{_\text{H}}\cdot A,   
\end{equation}
where $A$ is the cosmological gauge field $A_{\mu_1 \dots \mu_{D-1}}$, and the integration is taken over the horizon, which is a codimension-2 null hypersurface. In Eq. \eqref{Theta H}, one has $(\xi_\H\cdot A)_{\mu_1\dots \mu_{D-2}}=\xi_\H^\mu A_{\mu\mu_1\dots \mu_{D-2}}$.  This definition was inspired by the definition of the electric potential $\Phi_\H$ which was given above. So, it is appropriate to use the name \emph{cosmological potential} for $\Theta_{_\text{H}}$.

The new term $\Theta_{_\text{H}}\delta C$ in the first law resolves some issues related to the volume-pressure term $V\delta P$  which has been used before in the literature \cite{Teitelboim:1985dp,Henneaux:1989zc}(review \cite{Kubiznak:2016qmn}). Let us elaborate on this.
\begin{itemize}
\item $\Theta_{_\text{H}}$ is a property of the event horizon similar to the other horizon parameters $(T_{_\text{H}}, \Omega_{_\text{H}}, \Phi_{_\text{H}})$, in contrast with the volume $V$, which conceptually cannot be a property of the horizon, if it is  considered to be some volume inside the black hole. 
\item $\delta C$ is variation of a charge which is a parameter in the solution similar to $(M,S,J,Q)$ and in contrast with $\delta P$, which has been considered to be proportional to $\delta \Lambda$, i.e., variation of a parameter in the Lagrangian.
\item $\Theta_{_\text{H}}$ and $C$ are intensive and extensive quantities, respectively, and they are on the same foot as other terms in the first law. This is in contrast with $V \delta P$, where $V$ and $P$ are extensive and intensive, respectively. 
\item Noting the order of intensive and extensive quantities in $\Theta_\H\delta C$, the $M$ in the first law \eqref{first law} would be the energy or mass, in agreement with conserved charge being associated with time translation. This resolves the problem of promoting $M$ to be enthalpy \cite{Kastor:2009wy,Dolan:2010ha} (as a result of the inverse order of extensive or intensivity of $V\delta P$), which is inconsistent with $M$ as the conserved charge of the time translation symmetry. 
\item The conceptual problem with the \emph{negative} pressure for de Sitter spacetime is resolved, because the charge $C$, which is conceptually and mathematically similar to the electric charge, can be positive or negative.
\end{itemize}

In this paper, we continue the analysis in Ref. \cite{1} in three aspects: First, we revisit the Smarr formula in the presence of the cosmological charge $C$. Second, we show that the definition of $\Theta_{_\text{H}}$ in Eq. \eqref{Theta H} reproduces successfully an \emph{ad hoc} (but successful) volume term introduced in Ref.\cite{Kastor:2009wy} called effective volume. And finally, we fix a freedom or ambiguity in the definition of effective volume in the literature, which will be discussed in detail, by fixing the gauge freedom in the cosmological gauge field $A$ such that mass and other charge variations are reproduced correctly when the solution is perturbed by $\Lambda \to \Lambda+\delta \Lambda$.  The rest of the paper is devoted to a case study of different black hole solutions in different dimensions and theories. Using examples, we examine the reliability of the $\Theta_{_\text{H}}\delta C$ as a universal generalization of the first law. Besides, we enhance all black hole solutions by finding the cosmological gauge field $A_{\mu_1 \dots \mu_{D-1}}$ for them and presenting complete solutions as a reference for interested readers. We will also see that studying these examples sheds light on the universality of the Smarr formula for all $D\geq 3$ dimensions.

\section{Smarr formula and the effective volume}

\subsection{Smarr formula in the presence of a cosmological charge}\label{Smarr sec}
The first law of black hole thermodynamics is a universal constraint between the variations of conserved charges. It is universal, in the sense that it is independent of the spacetime dimension, theory, and the Lagrangian, asymptotic conditions, and the topology of the black hole. There is another constraint in black hole thermodynamics, a constraint between conserved charges (not their variations) which is called the Smarr relation \cite{Smarr:1972kt}. This relation is not a universal one. Especially, it explicitly depends on the dimensions of spacetime. Here, we show that, in the presence of a cosmological charge, this relation becomes
\begin{equation}\label{Smarr}
\boxed{(D-3) M=(D-2)T_{_\text{H}}S+(D-2)\Omega_{_\text{H}} J+ (D-3)\Phi_{_\text{H}} Q -\Theta_{_\text{H}} C. }
\end{equation}

In order to obtain this relation, we use the "scaling method" which is a well-known way to derive the Smarr formula (see, e.g., \cite{Townsend:1997ku,Kastor:2009wy}). In this method, $M$ is considered to be a homogeneous function of other charges $(S,J,Q,C)$. Using the Euler theorem, for a  function $f(p_1,p_2,\dots)$ homogeneous in the variables $(p_1, p_2, \dots )$ (in other words, for a constant $\alpha$, one has $\alpha^r f(p_1,p_2,...)=f(\alpha^{q_1}p_1, \alpha^{q_2}p_2, \dots)$), one can show that
\begin{equation}\label{Euler}
r f(p_1,p_2,\dots)=\sum_i q_i\left (\frac{\partial f}{\partial p_i}\right)p_i, \qquad i=1,2,\dots .    
\end{equation}
We can find the degree of homogeneity in $M=M(S,J,Q,C)$ ( {\it i.e.} the $r$ and $q_i$ in the above equation) using dimensional analysis.  Newton's constant is dimensionful and, hence, plays a role in the scaling of charges, but as a convenient convention, we set $G=1$ hereafter. By dimensional analysis, $M\sim l^{D-3}$, $S\sim l^{D-2}$,  $J\sim l^{D-2}$, $Q\sim l^{D-3}$, and $C\sim l^{-1}$, where $l$ is a length scale. Therefore, after scaling $l \to \alpha l$, one has
\begin{equation}\label{proof Smarr}
\alpha^{D-3}M(S,J,Q,C)=M\left(\alpha^{D-2}S,\alpha^{D-2}J,\alpha^{D-3}Q,\alpha^{-1}C\right).    
\end{equation}
Using Eqs. \eqref{Euler} and \eqref{proof Smarr}, one gets
\begin{equation}
(D-3) M=(D-2)\left(\frac{\partial M}{\partial S}\right)S+(D-2)\left(\frac{\partial M}{\partial J}\right) J+ (D-3)\left(\frac{\partial M}{\partial Q}\right) Q -\left(\frac{\partial M}{\partial C}\right)C. 
\end{equation}
Finally, using the first law \eqref{first law}, we find the Smarr relation \eqref{Smarr}. Needless to say, the analysis above is not a rigorous proof but only a heuristic justification. The Smarr relation may fail in some cases, especially if there are dimensionful quantities other than the conserved charges, as we shall see in some massive gravity theories below.

\subsection{Reproducing the effective volume}
Since the realization of $\Lambda$ as a pressure term in the first law, it has been a challenge how to find its thermodynamic conjugate, a ``volume" for a black hole. One way to circumvent this problem in the literature has been defining the volume by the first law itself, sometimes called ``thermodynamic volume." However, in Ref. \cite{Kastor:2009wy} an \emph{ad hoc} but successful (and, importantly, independent from the first law) definition for a viable black hole volume, called ``the effective volume"  was proposed. It is defined at the horizon by the formula
\begin{equation}\label{V eff}
V_{\text{eff}}\equiv \oint_\H \star \omega, \qquad \nabla_\mu\omega^{\mu\nu} \equiv \xi_\H^\nu.  
\end{equation}
Notice that $\omega$ is defined by the latter equation, {\it i.e.} $\xi_\H^\nu=\nabla_\mu\omega^{\mu\nu}$, and is ambiguous; one can deform it by $\omega \to \omega+\omega'$ with an arbitrary divergence-free term: $\nabla_\mu \omega'^{\mu\nu}=0$ (see examples in Ref. \cite{Cvetic:2010jb}). $\xi_\H^\nu$ is the Killing vector at the horizon, and the 2-form $\omega_{\mu\nu}$ is  called ``the Killing potential," and  its Hodge dual $\star \omega$ is a $(D-2)$-form which appears in the integrand of Eq. \eqref{V eff}. 

Here, we show how the potential $\Theta_\H$ in Eq. \eqref{Theta H} reproduces the $V_{\text{eff}}$ via the equation
\begin{equation}\label{Theta Veff}
\boxed{\Theta_\H=\pm \sqrt{|\Lambda|}  V_{\text{eff}}.}  
\end{equation}
To this end, we begin from the definition of $\Theta_\H$ in Eq. \eqref{Theta H}, denoting the Hodge dual of the integrand in it by $\tilde\omega$, {i.e.,}  
\begin{align}\label{omega t}
\star \tilde \omega\equiv \xi_\H\cdot A.    
\end{align}
By taking an exterior derivative of both sides,
\begin{align}
\mrd \star \tilde \omega&= \mrd (\xi_\H\cdot {A})\\
&= \mathcal{L}_{\xi_\H}{A}-\xi_\H\cdot \mrd {A}\\
&=-\xi_\H\cdot \mrd {A}. \label{V eff p1}
\end{align}
In the first equation, we used the Cartan identity $\mathcal{L}_{\xi}\boldsymbol{a}=\xi\cdot \mrd \boldsymbol{a}+\mrd (\xi\cdot \boldsymbol{a})$, which is correct for any differential form $\boldsymbol{a}$ and any vector field $\xi$. In the second equation, the isometry or Killing relation $\mathcal{L}_{\xi_\H}{A}=0$ was used. Using ${F}=\mrd{A}$, the on-shell relation \eqref{F on-shell}, and definition of the Hodge duality, we find from Eq. \eqref{V eff p1}
\begin{align}
\mrd \star \tilde \omega= \pm\sqrt{|\Lambda|}(\star \xi_\H).  
\end{align}
Applying the Hodge duality to both sides and using the identities $(\star \mrd \star \tilde \omega)^\nu=(-1)^D\nabla_\mu \tilde\omega^{\mu\nu}$ and $\star^2\xi_\H=(-1)^D\xi_\H$ (see Eqs. (A.19) and (A.29) in Appendix A in Ref. \cite{Hajian:2015eha}), then
\begin{equation}
\nabla_\mu \tilde\omega^{\mu\nu}=\pm\sqrt{|\Lambda|} \xi_\H^\nu.    
\end{equation}
Comparing this result with the  ``Killing potential" $\xi_\H^\nu=\nabla_\mu\omega^{\mu\nu}$ in Ref. \cite{Kastor:2009wy}, one finds 
\begin{equation}\label{omega t2}
\tilde\omega^{\mu\nu}=\pm\sqrt{|\Lambda|} \omega^{\mu\nu} \qquad \Rightarrow \qquad \oint_\text{H} \star\tilde\omega= \pm\sqrt{|\Lambda|}\oint_\text{H} \star\omega.   
\end{equation}
From this result and using Eqs.\eqref{Theta H}, \eqref{V eff}, and \eqref{omega t}, one arrives at the desired result:
\begin{equation}
\Theta_\H=\pm \sqrt{|\Lambda|}  V_{\text{eff}}.  \nonumber    
\end{equation}

An astute reader might wonder about the extra factor $\pm \sqrt{|\Lambda|}$ appearing in the equation above.  This factor is not unexpected because the charge $C$ and $\Lambda$ are quadratically related in Eq. \eqref{C}, so 
\begin{equation}\label{del C}
\delta C= \frac{\pm \delta \Lambda}{8\pi\sqrt{|\Lambda|}}.   
\end{equation}
By the relation $\frac{ \delta \Lambda}{8\pi}\equiv \delta P$, we realize that $\delta C= \frac{\pm \delta P}{\sqrt{|\Lambda|}}$. So, the extra factor $\pm \sqrt{|\Lambda|}$  in $\Theta_\H=\pm \sqrt{|\Lambda|}  V_{\text{eff}}$ is canceled with the extra factor $\frac{\pm 1}{\sqrt{|\Lambda|}}$ in $\delta C$, yielding same final result, i.e.,  $\Theta_{_\text{H}}\delta C=V_{\text{eff}} \delta P$.

\subsection{Fixing the gauge freedom}
As was mentioned in the previous section, the effective volume has an ambiguity: a  divergence-free 2-form can be added to the Killing potential 
\begin{equation}
\omega_{\mu\nu} \to \omega_{\mu\nu}+\omega'_{\mu\nu}, \qquad \nabla_\mu \omega'^{\mu\nu}=0.    
\end{equation}
Using the $\tilde \omega$ to relate Eqs. \eqref{omega t} and \eqref{omega t2}, it is easy to see that this ambiguity is related to the gauge freedom in $A\to A+\mrd \lambda$ as
\begin{equation}
\omega'=\star\left( \frac{\xi_\H\cdot \mrd \lambda}{\pm \Lambda}\right).    
\end{equation}
As a result, one can fix $\lambda$ in the ``charge formulation of $\Lambda$" in order to remove the $\omega'$ ambiguity in the definition effective volume. To this end, we notice that the cosmological gauge field and its variations appear explicitly in the covariant formulation of charges (see Appendix \ref{appendix cov}). In order to reproduce the variations of mass, angular momentum, and other conserved charges with respect to $\delta \Lambda$, the gauge fixing plays an important role, as we will clarify this issue with different examples.   

Summarizing this section, we generalized the Smarr relation to include the contribution from the cosmological conserved charge $\Theta_\H C$. Moreover, it was shown how $\Theta_{_\text{H}}\delta C$ in the cosmological charge formulation reproduces $V_{\text{eff}} \delta P$, while resolving its conceptual and computational problems as well as removing its ambiguity by gauge fixing. In particular, the $\Theta_\H$ reproduces the $V_{\text{eff}}$ as the potential associated with the gauge field $A$ on the horizon. In the rest of the paper, we examine the power of this formulation by studying different examples explicitly. Importantly, we provide the cosmological gauge field $A_{\mu_1 \dots \mu_{D-1}}$  and the corresponding black hole cosmological potential $\Theta_{_\text{H}}$ for all of these black hole and brane solutions, and check the first law and the Smarr relation for all of them. 

\section{Examples: solutions in three dimensions}
We start our analysis of explicit solutions in three dimensions with the simplest example, the Banados-Teitelboim-Zanelli (BTZ) black hole \cite{Banados:1992wn}. We give the details of the calculations for the BTZ black hole, but we will give only the results of the computations for other examples to avoid repetition.   
\subsection{BTZ black hole in the cosmological Einstein gravity}\label{BTZ sec}
\subsubsection{Theory}  Einstein-$\Lambda$ theory in $3$-dim
\begin{equation}
\mathcal{L}=\frac{1}{16\pi }(R-2\Lambda).
\end{equation}
\subsubsection{Solution} The metric in the coordinates $x^\mu=(t,r,\varphi)$ is \cite{Banados:1992wn} 
\begin{align}
&\mrd s^2= -\Delta\mrd t^2 +\frac{\mrd r^2}{\Delta}+r^2(\mrd \varphi-\omega \mrd t)^2, \qquad \Delta\equiv -m+\frac{r^2}{\ell^2}+\frac{j^2}{4r^2}, \qquad \omega\equiv \frac{j}{2r^2},
\end{align}
where $\Lambda=\dfrac{-1}{\ell^2}$. The outer and inner horizons are located at $2r_\pm^2=\ell^2 (m\pm\sqrt{m^2-\frac{j^2}{\ell^2}})$. The cosmological gauge field for this black hole solution can be found to be (see Appendix \ref{app A} for more details)
\begin{equation}\label{A example 3-1}
    A = -\frac{r^2}{2\ell} \mrd t \wedge \mrd \varphi. 
\end{equation}
Notice that one can add a term $f(m,j,\ell)\mrd t\wedge \mrd \varphi$ to $A$, which clearly does not change the field strength $F=\mrd A$ if $f$ is not a function of spacetime coordinates. This is a simple example of the gauge freedom that we have discussed before. Nonetheless, the $A$ and its variations with respect to $m,j,\ell$ appear explicitly in the charge calculations (mass, angular momentum and cosmological charge). To see this, Appendix-\ref{appendix cov} is provided.  Requesting the charges to be reproduced correctly in the new paradigm (in comparison to the  usual paradigm of $\Lambda$ being a constant in Lagrangian) fixes the gauge freedom for our example to be what it is already in Eq.\eqref{A example 3-1}.   

\subsubsection{Properties} 
\begin{align}\label{Prop 3.1}
 M=\frac{m}{8}, \qquad  J=\frac{j}{8},\qquad \Omega_\pm=\frac{r_\mp}{\ell r_\pm}, \qquad T_\pm=\frac{r_\pm^2-r_\mp^2}{2\pi \ell^2 r_\pm}, \qquad S_\pm=\frac{\pi r_\pm}{2}.
\end{align}
The horizon Killing vectors are $\xi_\pm=\partial_t+\Omega_\pm \partial_\varphi$. 
Using $A$ from Eq. \eqref{A example 3-1} in the definition of $\Theta_{_\text{H}}$ in Eq. \eqref{Theta H}, we get
\begin{align}
    \Theta_\pm &= \int_{r_\pm} \left( \partial_t + \Omega_\pm \partial_{\phi} \right) \cdot \left( \frac{-r^2}{2\ell} \mrd t \wedge \mrd \phi \right)\\
    &=\int_{r_\pm} \frac{-r^2}{2\ell}\mrd \varphi - \int_{r_\pm} \Omega_\pm \frac{-r^2}{2\ell} \mrd t.
\end{align}
However, the last integral vanishes because the pullback of the $\mrd t$ to the surface of integration (which is the bifurcation point of the horizon parametrized by the coordinate $\varphi$) vanishes. So,
\begin{align}\label{Theta 3.1}
    \Theta_\pm =\int_{r_\pm} \frac{-r^2}{2\ell}\mrd \varphi =-\frac{\pi r_\pm^2}{\ell}.
\end{align}
The cosmological charge $C$ can be read from Eq. \eqref{C} with the lower sign (which is the one for negative $\Lambda$) to be 
\begin{equation}
C=-\frac{1}{4\pi \ell}.    
\end{equation}

\subsubsection{The first law and the Smarr formula}
The generalized first law and the Smarr formula for the BTZ black hole read
\begin{align}
&\delta M=T_\pm \delta S_\pm+\Omega_\pm \delta J+\Theta_\pm \delta C, \label{First law 3.1}\\
&0 = T_\pm S_\pm + \Omega_\pm J - \Theta_\pm C \label{Smarr 3.1}
\end{align}
respectively. One can check the validity of these two relations explicitly which we do next. Let us check the Smarr formula first. Substituting Eqs. \eqref{Prop 3.1} and \eqref{Theta 3.1} to Eq. \eqref{Smarr 3.1}, one has 
\begin{equation}
    0 =  \frac{r_\pm^2-r_\mp^2}{2\pi \ell^2 r_\pm} \frac{\pi r_{\pm}}{2} + \frac{r_{\mp}}{\ell r_{\pm}} \frac{j}{8} - \frac{-\pi r^2_{\pm}}{\ell}\frac{-1}{4\pi \ell} \qquad \Rightarrow \qquad   0=r_\mp\left (\frac{- r_{\mp}}{4 \ell^2} + \frac{j}{8\ell r_{\pm}}\right), 
\end{equation}
which is satisfied for $2r_\pm^2=\ell^2 \left(m\pm\sqrt{m^2-\frac{j^2}{\ell^2}}\right)$. Hence, the Smarr formula holds.

Now, let us look at the validity of the first law of black hole thermodynamics \eqref{First law 3.1}. This solution has three independent parameters $m$, $j$, and $\ell$. Notice that $\ell$ is a free parameter of the solution, if the Lagrangian \eqref{L F2} is the Lagrangian describing the theory of gravity. In other words, $\ell$ is related to $\Lambda$ by $\Lambda=\frac{-1}{\ell^2}$, and $\Lambda$ is related to $c$ in Eq. \eqref{F on-shell} (which is a free parameter of the solution), by the relation \eqref{EoM Lambda}. We calculate variations to nearby black hole solutions with respect to each of these three parameters. This method of variation can be called parametric variations \cite{Hajian:2014twa}. We can begin with variation in the $m$ parameter:
\begin{align}\label{3.1 a}
  \delta_m M = \frac{\delta m}{8}, \qquad  \delta_{m} S_{\pm} &=\frac{\pi}{2} \delta_{m} r_{\pm}, \qquad \delta_{m} J =0, \qquad \delta_m C = 0,
  \end{align}
where $\delta_{m} r_{\pm}$ reads as follows:
\begin{align}\label{3.1 b}
\delta_m r_\pm&=\frac{\partial r_\pm}{\partial m}\delta m=\pm\frac{\sqrt{2\ell^3(m\ell\pm\sqrt{m^2\ell^2-j^2})}}{4\sqrt{m^2\ell^2-j^2}}\delta m.
\end{align}
Substituting Eqs. \eqref{3.1 a} and \eqref{3.1 b} in the first law \eqref{First law 3.1}, one finds
\begin{align}
\frac{\delta m}{8}= \left(\frac{r_\pm^2-r_\mp^2}{2\pi \ell^2 r_\pm}\right)\left(\pm\frac{\pi}{2}\frac{\sqrt{2\ell^3(m\ell\pm\sqrt{m^2\ell^2-j^2})}}{4\sqrt{m^2\ell^2-j^2}}\delta m\right)\,\Rightarrow\, \frac{r_\pm}{r_\pm^2-r_\mp^2}=\pm\frac{\sqrt{\frac{\ell^2}{2}(m\pm\sqrt{m^2-\frac{j^2}{\ell^2}})}}{\ell^2\sqrt{m^2-\frac{j^2}{\ell^2}}},   
\end{align}
which is satisfied by $r_\pm$. Similarly, for the variation of $j$,
\begin{align}\label{3.1 c}
  \delta_j M = 0, \qquad  \delta_{m} S_{\pm} &=\frac{\pi}{2} \delta_{j} r_{\pm}, \qquad \delta_{j} J =\frac{\delta j}{8}, \qquad \delta_j C = 0,
  \end{align}
in which 
\begin{equation}
\delta_j r_\pm= \mp \frac{\sqrt{2\ell}\, j\delta j}{4\sqrt{m^2\ell^2-j^2}(m\ell-\sqrt{m^2\ell^2-j^2})}.    
\end{equation}
Inserting in the first law \eqref{First law 3.1}, we find
\begin{align}
&0=\left(\frac{r_\pm^2-r_\mp^2}{2\pi \ell^2 r_\pm}\right)\left(\mp\frac{\pi}{2}\frac{\sqrt{2\ell}\, j\delta j}{4\sqrt{m^2\ell^2-j^2}(m\ell\pm\sqrt{m^2\ell^2-j^2})}\right)+\frac{r_\mp}{\ell r_\pm}\frac{\delta j}{8} \\
\Rightarrow\qquad &\frac{r_\mp}{r_\pm^2-r_\mp^2}=\pm\frac{\sqrt{2\ell}\, j}{4\ell(\sqrt{m^2\ell^2-j^2})\sqrt{m\ell\pm\sqrt{m^2\ell^2-j^2}}}.\label{3.1 d}
\end{align}
Using the relations $\pm\ell\sqrt{m^2\ell^2-j^2}=(r_\pm^2-r_\mp^2)$ and $\ell\sqrt{m\pm\sqrt{m^2-j^2/\ell^2}}=\sqrt{2}r_\pm$ in the denominator of the right-hand side, the result in Eq. \eqref{3.1 d} simplifies to
$4r_\pm r_\mp=2\ell j$, which is the correct equation, admitting the first law to be satisfied. 

We should also check the first law for the variation with respect to $\ell$. To this end we have
\begin{align}\label{3.1 e}
  \delta_\ell M = 0, \qquad  \delta_{\ell} S_{\pm} &=\frac{\pi}{2} \delta_{\ell} r_{\pm}, \qquad \delta_{\ell} J =0, \qquad \delta_\ell C = \frac{\delta \ell}{4\pi \ell^2},
  \end{align}
in which 
\begin{equation}
\delta_\ell r_\pm= \pm \frac{(2m^2\ell^2-j^2\pm 2m\ell\sqrt{m^2\ell^2-j^2})\delta \ell}{4\sqrt{\frac{\ell}{2}(m^2\ell^2-j^2)(m\ell\pm\sqrt{m^2\ell^2-j^2})}}.    
\end{equation}
Putting these in the first law \eqref{First law 3.1}, it follows that
\begin{align}
& 0=(\frac{r_\pm^2-r_\mp^2}{2\pi \ell^2 r_\pm})\left(\pm \frac{\pi}{2}\frac{(2m^2\ell^2-j^2 \pm 2m\ell\sqrt{m^2\ell^2-j^2})\delta \ell}{4\sqrt{\frac{\ell}{2}(m^2\ell^2-j^2)(m\ell\pm\sqrt{m^2\ell^2-j^2})}}\right)+(-\frac{\pi r_\pm^2}{\ell})(\frac{\delta \ell}{4\pi \ell^2})\\
\Rightarrow \quad & \frac{r^3_\pm}{r_\pm^2-r_\mp^2}=\pm\frac{\ell(2m^2\ell^2-j^2 \pm 2m\ell\sqrt{m^2\ell^2-j^2})}{4\sqrt{\frac{\ell}{2}(m^2\ell^2-j^2)(m\ell\pm\sqrt{m^2\ell^2-j^2})}}.
\end{align}
Using the relations $\pm\ell\sqrt{m^2\ell^2-j^2}=(r_\pm^2-r_\mp^2)$ and $\ell\sqrt{m\pm\sqrt{m^2-j^2/\ell^2}}=\sqrt{2}r_\pm$ in the denominator of the right-hand side, it reduces to $4r^4_\pm=\ell^2(2m^2\ell^2-j^2 \pm 2m\ell\sqrt{m^2\ell^2-j^2})$, which is the correct equation. 

According to the analysis above, we deduce that the generalized first law in Eq. \eqref{First law 3.1} and the Smarr formula in Eq. \eqref{Smarr 3.1}, which include the new terms $\Theta_\H\delta C$ and $\Theta_\H C$,  are correct relations for this example and confirm the results of the analysis in this paper.

\subsection{Charged static BTZ black hole}
\subsubsection{Theory}  Einstein-Maxwell-$\Lambda$ theory in $2+1$ dimensions
\begin{equation}
\mathcal{L}=\frac{1}{16\pi }(R-2\Lambda-F_{\mu\nu}F^{\mu\nu}).
\end{equation}
\subsubsection{Solution} The metric and the Maxwell gauge field in the coordinates $x^\mu=(t,r,\varphi)$ are \cite{Martinez:1999qi} 
\begin{align}
&\mrd s^2= -\Delta\mrd t^2 +\frac{\mrd r^2}{\Delta}+r^2\mrd \varphi^2, \qquad \Delta\equiv -m+\frac{r^2}{\ell^2}-\frac{q^2}{2}\log{\frac{r}{\ell}},\\
&A=-\frac{q}{2}\log (\frac{r}{\ell})\, \mrd t,\nonumber
\end{align}
with $\Lambda=\dfrac{-1}{\ell^2}$. Horizons are at $\Delta=0$. For this solution, the cosmological gauge field $\boldsymbol{A}$ (denoted bold in order to be distinguished from the Maxwell field $A_\mu$) in an appropriate gauge is (see Appendix \ref{app A})
 \begin{equation}\label{A example 3-2}
\boldsymbol{A}=-(\frac{4r^2-q^2\ell^2}{8\ell})\mrd t\wedge \mrd \varphi.     
 \end{equation}
The gauge freedom of $\boldsymbol{A}$ is fixed such that it reproduces the variation of the mass and the other charges with respect to $\ell$ correctly. To see this, one can use the covariant phase space formulation of charges. The details of the formulation are described in Refs. \cite{ HS:2015xlp, Ghodrati:2016vvf, 1}. However, for the sake of completeness, we have added Appendix \ref{appendix cov} which provides the final formula to perform such charge calculations. 

\subsubsection{Properties} 
Horizon Killing vectors are $\xi_\H=\partial_t$. 
Using $\boldsymbol{A}$ from Eq. \eqref{A example 3-2} in the definition of $\Theta_{_\text{H}}$ in Eq.  \eqref{Theta H}, we get
\begin{align}
    \Theta_\H &= \int_{r_\H} \left( \partial_t \right) \cdot \left( \frac{-(4r^2-q^2\ell^2)}{8\ell} \mrd t \wedge \mrd \varphi \right)=\int_{r_\H} (\frac{-(4r^2-q^2\ell^2)}{8\ell})\mrd \varphi=-\frac{\pi(4r_\H^2-q^2\ell^2)}{4\ell}.
\end{align}
For the other properties, including $C$ from Eq. \eqref{C}, we find
\begin{align}
&M=\frac{m}{8}, \qquad Q=\frac{q}{4}, \qquad C=-\frac{1}{4\pi \ell}, \qquad  \Theta_\H =-\frac{\pi(4r_\H^2-q^2\ell^2)}{4\ell}\nonumber\\
&\Phi_\H=-\frac{q}{2}\log{\frac{r_\H}{\ell}}, \qquad T_\H=\frac{4r_\H^2-q^2\ell^2}{8\pi r_\H\ell^2}, \qquad S_\H=\frac{\pi r_\H}{2}. \label{Prop 3.2}
\end{align}
The generator of the entropy as a conserved charge is $\eta_{_\text{H}}=\frac{1}{T_{_\text{H}}}\{\partial_t,-\Phi_{_\text{H}}\}$ \cite{HS:2015xlp,Ghodrati:2016vvf}.

\subsubsection{The first law and the Smarr formula} 
The generalized first law and the Smarr formula for this solution are
\begin{align}
&\delta M=T_\H \delta S_\H+\Phi_\H \delta Q+\Theta_\H \delta C, \label{First law 3.2}\\
&0 = T_\H S_\H - \Theta_\H C \label{Smarr 3.2}
\end{align}
respectively. The Smarr relation can be checked easily using Eq. \eqref{Prop 3.2}. To check the first law,  notice that the solution has three free parameters $m$,$q$, and $\ell$. Using the relations
\begin{align}
\delta_m r_\H= \frac{2\ell^2 r_\H \delta m}{4r_\H^2-q^2\ell^2}, \qquad  \delta_q r_\H=\frac{2q\ell^2 r_\H \log (\frac{r_\H}{\ell})\delta q}{4r_\H^2-q^2\ell^2}, \qquad   \delta_\ell r_\H=\frac{r_\H}{\ell}\delta \ell,
\end{align}
and following the same steps as in Sec. 3.1, the first law can also be checked. The result is affirmative, and the first law holds for the charged static BTZ black hole.

\subsection{Lifshitz $z=3$ black hole }
\subsubsection{Theory}  New Massive Gravity (NMG) theory in $2+1$ dimensions \cite{Bergshoeff:2009hq}
\begin{equation}\label{NMG Lagrangian}
\mathcal{L}=\frac{1}{16\pi }\left(R-2\Lambda+\frac{1}{\mathfrak{m}^2}(R_{\mu\nu}R^{\mu\nu}-\frac{3}{8}R^2)\right).
\end{equation}
\subsubsection{Solution} The metric in the coordinates $x^\mu=(t,r,\varphi)$ is \cite{AyonBeato:2009nh,AyonBeato:2010tm}
\begin{equation}
\mrd s^2=-(\frac{r}{\ell})^{2z}(1-\frac{m\ell^2}{r^2})\,\mrd t^2 +\frac{\mrd r^2}{\frac{r^2}{\ell^2}(1-\frac{m\ell^2}{r^2})}+r^2 \mrd \varphi^2
 \end{equation} 
for $z=3$, one has $\Lambda=-\frac{13}{2\ell^2}$ and $\mathfrak{m}^2=\frac{1}{2\ell^2}$. Notice that $\mathfrak{m}$ and $m$ are different parameters: The former is a parameter in the Lagrangian, and the latter is a parameter of the solution. The event horizon is at $r_\H=\sqrt{m\ell^2}$. The cosmological gauge field in an appropriate gauge for this solution is (see Appendix \ref{app A})
\begin{equation}\label{A example 3-3}
A=\sqrt{|\Lambda|}\left(\frac{3m^2}{8\Lambda}-\frac{r^4}{4\ell^2}\right)\mrd t\wedge \mrd \varphi.    
\end{equation}

\subsubsection{Properties}
For this black hole, one can find  \cite{Hohm:2010jc,Gonzalez:2011nz,Gim:2014nba,Ayon-Beato:2015jga,Ghodrati:2016vvf}
\begin{align}\label{Prop 3.3}
&M=\frac{m^2}{4}, \qquad C=-\sqrt{\frac{13}{2}}\frac{1}{4\pi \ell}, \qquad  T_\H=\frac{r_\H^3}{2\pi\ell^4}, \qquad  {S}_\H={2\pi}r_\H.
\end{align}
Using the horizon Killing vector $\xi_\H=\partial_t$ and $A$ from Eq. \eqref{A example 3-3} in Eq. \eqref{Theta H}, we find
\begin{align}\label{Theta 3.3}
    \Theta_\H &= \int_{r_\H}\! \! \left( \partial_t \right) \cdot \left( \sqrt{|\Lambda|}(\frac{3m^2}{8\Lambda}-\frac{r^4}{4\ell^2}) \mrd t \wedge \mrd \varphi \right)=\int_{r_\H}  \!\! \sqrt{|\Lambda|}(\frac{3m^2}{8\Lambda}-\frac{r^4}{4\ell^2}) \mrd \varphi = -\sqrt{\frac{2}{13}}4\pi m^2\ell,
\end{align}
where, in the last equality, we used $\Lambda=-\frac{13}{2\ell^2}$ and $r_\H=\sqrt{m\ell^2}$.

\subsubsection{The First law and the Smarr formula} 
This solution has two free parameters $m$ and $\ell$. The horizon radius in terms of these two parameters is $r_\H=\sqrt{m\ell^2}$, which makes the calculations very simple. Using Eqs. \eqref{Prop 3.3} and \eqref{Theta 3.3}, the generalized first law and Smarr formula for this solution 
\begin{align}
&\delta M=T_\H \delta S_\H+\Theta_\H \delta C, \label{First law 3.3}\\
&0 = T_\H S_\H - \Theta_\H C \label{Smarr 3.3}
\end{align}
respectively, can be checked easily for variations with respect to $m$ and $\ell$. Hence, for this example the first law and Smarr formula hold.

\subsection{BTZ black hole in the New Massive Gravity}

\subsubsection{Theory} The theory is again the NMG theory in three dimensions \cite{Bergshoeff:2009hq}:
\begin{equation}
\mathcal{L}=\frac{1}{16\pi }\left(R-2\Lambda+\frac{1}{\mathfrak{m}^2}(R_{\mu\nu}R^{\mu\nu}-\frac{3}{8}R^2)\right).
\end{equation}
\subsubsection{Solution} The solution is exactly the same BTZ solution reviewed in Sec. \ref{BTZ sec}, i.e.,  in coordinates $x^\mu=(t,r,\varphi)$, the metric is \cite{Banados:1992wn,Clement:2009gq} 
\begin{align}
&\mrd s^2= -\Delta\mrd t^2 +\frac{\mrd r^2}{\Delta}+r^2(\mrd \varphi-\omega \mrd t)^2, \qquad \Delta\equiv -m+\frac{r^2}{\ell^2}+\frac{j^2}{4r^2}, \qquad \omega\equiv \frac{j}{2r^2},
\end{align}
 but for $\Lambda=\frac{-1}{\ell^2}+\frac{1}{4\ell^4\mathfrak{m}^2}$ and we assume also $\Lambda<0$.   Horizons are at $2r_\pm^2=\ell^2 (m\pm\sqrt{m^2-\frac{j^2}{\ell^2}})$. Cosmological gauge field for this black hole solution can be found to be (see Appendix \ref{app A})
\begin{equation}\label{A BTZ-NMG}
    A = -\sqrt{|\Lambda|}\left(\frac{r^2}{2}-\frac{m\ell^2}{2(1-2\mathfrak{m}^2\ell^2)}\right) \mrd t \wedge \mrd \varphi. 
\end{equation}
The gauge freedom (i.e., the second term in the parentheses) is fixed such that using the covariant phase space formulation of charges (see Appendix \ref{appendix cov}), or other methods such as the Abbott-Deser-Tekin (ADT) formulation  \cite{Abbott:1981ff,Deser:2002rt,Deser:2002jk} yields correct mass variations with respect to $\ell$ as well as other solution parameters. 

\subsubsection{Properties} 
Although this black hole is exactly the same as the BTZ black hole in Sec. \ref{BTZ sec}, it is the solution to a different theory which affects the charges $M$, $J$, and $S$ \cite{Clement:2009gq,Alkac:2012bz}:
\begin{align}\label{Prop BTZ-NMG}
& M=(1+\frac{1}{2\ell^2\mathfrak{m}^2})\frac{m}{8}, \qquad  J=(1+\frac{1}{2\ell^2\mathfrak{m}^2})\frac{j}{8},\nonumber\\
&\Omega_\pm=\frac{r_\mp}{\ell r_\pm}, \qquad T_\pm=\frac{r_\pm^2-r_\mp^2}{2\pi \ell^2 r_\pm}, \qquad S_\pm=(1+\frac{1}{2\ell^2\mathfrak{m}^2})\frac{\pi r_\pm}{2}.
\end{align}
Horizon Killing vectors are $\xi_\pm=\partial_t+\Omega_\pm \partial_\varphi$. The cosmological charge and horizon potential for this solution are, respectively,
\begin{equation}
C=-\frac{\sqrt{|\Lambda|}}{4\pi}, \qquad \Theta_\pm= -\pi \sqrt{|\Lambda|}\left(r_\pm^2-\frac{m\ell^2}{1-2\mathfrak{m}^2\ell^2}\right).
\end{equation}

\subsubsection{The first law and the Smarr formula}
The generalized first law for this solution is
\begin{align}
&\delta M=T_\pm \delta S_\pm+\Omega_\pm \delta J+\Theta_\pm \delta C,
\end{align}
which can be checked to be a correct relation by using variations with respect to three free parameters of this solution $m$, $j$, and $\ell$. On the other hand, the generalized Smarr formula is \emph{not} satisfied for this solution. To satisfy the Smarr formula, one needs to take into account the dimensionful quantity $\mathfrak{m}$ in a suitable way, which we have not been able to do so far.

\subsection{Horndeski BTZ-like black hole}
\subsubsection{Theory}  A Horndeski gravity in three dimensions \cite{Horndeski:1974wa} has the Lagrangian
\begin{equation}\label{Horndeski L}
\mathcal{L}=\frac{1}{16\pi}\Big(R-2\Lambda-2(\alpha g_{\mu\nu}-\gamma G_{\mu\nu})\nabla^\mu\phi\nabla^\nu\phi\Big),
\end{equation}
where $G_{\mu\nu}=R_{\mu\nu}-\frac{1}{2}Rg_{\mu\nu}$ is the Einstein tensor. 

\subsubsection{Solution} The metric in the coordinates $x^\mu=(t,r,\varphi)$ is \cite{Santos:2020xox}
\begin{align}
& \mrd s^2=-f \mrd t^2+\frac{\mrd r^2}{f}+r^2 (\mrd\varphi^2-\frac{j}{r^2}\mrd t), \nonumber\\ 
& f=-m+\frac{\alpha r^2}{\gamma}+\frac{j^2}{r^2}, \quad  \mrd\phi=\sqrt{\frac{-(\alpha+\gamma\Lambda)}{2\alpha \gamma f}}  \mrd r,
\end{align}
where $\gamma <0$ and $(m,j)$ are free parameters of the solution. The cosmological gauge field for this solution can be found to be (see Appendix \ref{app A})
\begin{equation}\label{A BTZ Horndeski}
A=-\sqrt{|\Lambda|}\left(\frac{r^2}{2}-\frac{\gamma m}{4\alpha}\right)\mrd t \wedge \mrd \varphi.    
\end{equation}
The gauge is fixed such that the covariant formulation of conserved charges (see Appendix \ref{appendix cov}) produces correct mass variation with respect to $\Lambda$, i.e., the $\delta_\Lambda M$. 

\subsubsection{Properties}
For this solution, the charges and the chemical potentials are computed to be \cite{Hajian:2020dcq}
\begin{align}
&M=\frac{(\alpha -\Lambda \gamma)m}{16 \alpha}, \qquad J=   \frac{(\alpha -\Lambda \gamma)j}{8 \alpha},\qquad   r_\pm^2=\frac{\gamma m \mp \sqrt{\gamma^2m^2-4\gamma\alpha j^2}}{2\alpha}, \nonumber\\
& \Omega_\pm=\frac{j}{r_\pm^2}, \qquad \kappa_\pm=\frac{\alpha(r^2_+-r^2_-)}{\gamma r_\pm}, \qquad T_\pm=\left(\frac{\alpha - \Lambda\gamma}{4\pi\alpha}\right) \kappa_\pm, \qquad S_\pm=\frac{\pi r_\pm}{2}.
\end{align}
Notice that $\alpha<0$ in order to have finite and positive horizon radii. Note also that the temperature is different from the usual  $\frac{\kappa}{2\pi}$ (i.e. the standard Hawking temperature) by a factor of  $\frac{\alpha - \Lambda\gamma}{4\pi\alpha}$, which is a result of the fact that, in Horndeski gravities, the effective speed of the graviton can be (as in our example here) different from $1$ \cite{Hajian:2020dcq}. The cosmological charge and the horizon potential, using Eqs. \eqref{Theta H} and \eqref{A BTZ Horndeski}, are, respectively, 
\begin{equation}
C=-\frac{\sqrt{|\Lambda|}}{4\pi}, \qquad \Theta_\pm=-\pi\sqrt{|\Lambda|}(r_\pm^2-\frac{\gamma m}{2\alpha}).   \end{equation}

\subsubsection{The first law and the Smarr formula} 
This solution has three free parameters $m$, $j$ and $\Lambda$. The generalized first law for this solution is
\begin{align}
&\delta M=T_\pm \delta S_\pm+\Omega_\pm \delta J+\Theta_\pm \delta C,
\end{align}
which can be checked to be a correct relation by using variations with respect to three free parameters of this solution. For this solution, the generalized Smarr formula is \emph{not} satisfied as in the previous example. So one should find the correct formula taking into account all the dimensionful parameters in the theory.

\section{Solutions in four dimensions}

\subsection{(A)dS-Kerr-Newman black hole} \label{Kerr Newman sec}
\subsubsection{Theory}  Einstein-Maxwell-$\Lambda$ theory in four dimensions \cite{Kerr:1963ud,Newman:1965tw,Newman:1965my,Carter:1968ks,Carter:1970ea,Carter:1970ea2}
\begin{equation}
\mathcal{L}=\frac{1}{16\pi }(R-2\Lambda-F_{\mu\nu}F^{\mu\nu}).
\end{equation}

\subsubsection{Solution} 
The metric in the coordinates $x^\mu=(t,r,\theta,\varphi)$ is
\begin{align}
\mathrm{d}s^2= -&\Delta_\theta(\frac{1-\frac{\Lambda r^2}{3}}{\Xi}-\Delta_\theta f)\mathrm{d}t^2+\frac{\rho ^2}{\Delta_r}\mathrm{d}r^2+\frac{\rho ^2}{\Delta_\theta} \mathrm{d}\theta ^2 -2\Delta_\theta fa\sin ^2 \theta\,\mathrm{d}t \mathrm{d}\varphi\nonumber \\
+&\left( \frac{r^2+a^2}{\Xi}+fa^2\sin ^2\theta \right)\sin ^2\theta\,\mathrm{d}\varphi ^2\,,
\end{align}
where
\begin{align}
\rho^2 &\equiv r^2+a^2 \cos^2 \theta\,,\qquad \Delta_r \equiv (r^2+a^2)(1-\frac{\Lambda r^2}{3})-2mr + q^2\,,\nonumber\\
\Delta_\theta&\equiv 1+\frac{\Lambda a^2}{3}\cos ^2\theta\,,\qquad \Xi\equiv 1+\frac{\Lambda a^2}{3}\,,\qquad
f\equiv\frac{2mr-q^2}{\rho ^2\Xi^2}\,.\nonumber
\end{align}
 In these coordinates, the Maxwell gauge field is
\begin{equation}
A=\frac{qr}{\rho^2\Xi}(\Delta_\theta\mrd t-a\sin^2 \theta \,\mrd \varphi)\,. 
\end{equation} 
For positive and negative signs of $\Lambda$, the solution is a de Sitter or anti-de Sitter Kerr-Newman black hole, respectively. The analysis here is independent of this sign, and we leave it to be either positive or negative. We denote the cosmological gauge field by $\boldsymbol{A}$, in order to distinguish it from the Maxwell gauge field $A$. For this solution, $\boldsymbol{A}$ can be found to be (see Appendix \ref{app A})
\begin{equation}\label{A Kerr-Newman}
\boldsymbol{A}=-\frac{\sqrt{|\Lambda|}(r^3+3ra^2\cos^2\theta+\frac{ma^2}{\Xi})\sin\theta}{3\Xi}\mrd t \wedge\mrd \theta \wedge\mrd \varphi.    
\end{equation}
Similar to the other solutions described above, the gauge is fixed if one demands that the mass, angular momentum, and other charges be reproduced correctly by the covariant formulation of charges.

\subsubsection{Properties}
One can find the thermodynamic variables for this solution irrespective of the sign of $\Lambda$ as \cite{Gibbons:1977mu,Hajian:2016kxx} 
\begin{align}\label{Prop 4.1}
&M=\frac{m}{\Xi^2}\,,\qquad J=\frac{ma}{\Xi^2}\,, \qquad Q=\frac{q}{\Xi}\,, \qquad \Phi_\H=\frac{qr_\H}{r_\H^2+a^2}\,,\nonumber \\
&\Omega_\H=\frac{a(1-\frac{\Lambda r_{_\mathrm H}^2}{3})}{r_{_\mathrm H}^2+a^2}, \qquad T_\H=\frac{r_{_\mathrm H}(1-\frac{\Lambda a^2}{3}-\Lambda{r_{_\mathrm H}^2}-\frac{a^2}{r_{_\mathrm H}^2})}{4\pi(r_{_\mathrm H}^2+a^2)},  \qquad {S_{_\mathrm H}}=\frac{\pi(r_{_\mathrm{H}}^2+a^2)}{\Xi},
\end{align}
in which $r_{_{\mathrm{H}}}$ is the radius of the considered horizon. The cosmological charge and potential can also be found by Eqs. \eqref{C} and \eqref{Theta H}:
\begin{equation} \label{Theta 4.1}
C=\pm\frac{\sqrt{|\Lambda|}}{4\pi}, \qquad \Theta_\H=-\frac{\sqrt{|\Lambda|}4\pi(r_\H^3+r_\H a^2+\frac{ma^2}{\Xi})}{3\Xi}.  
\end{equation}
The upper and lower signs are for de Sitter and anti-de Sitter black holes, respectively. 

\subsubsection{The first law and the Smarr formula}
This solution has four free parameters $(m,a,q,\Lambda)$.  Using Eqs. \eqref{Prop 4.1} and \eqref{Theta 4.1}, the generalized first law and Smarr formula for this solution 
\begin{align}
&\delta M=T_\H \delta S_\H+\Omega_\H\delta J+\Phi_\H\delta Q+\Theta_\H \delta C, \label{First law 4.1}\\
&M = 2T_\H S_\H +2\Omega_\H J +\Phi_\H Q- \Theta_\H C, \label{Smarr 4.1}
\end{align}
respectively, can be checked for variations with respect to the parameters $p_i\in \{m,a,q,\Lambda\}$. Hence, for this example the generalized first law and Smarr formula hold.  In Appendix \ref{app first law Smarr},  the methods of checking the first law and Smarr formula are described, if the horizon radii cannot be found explicitly in terms of the parameters $p_i$ of the solution.

\subsection{A black hole in Horndeski gravity}
\subsubsection{Theory}
The Lagrangian of the theory is \cite{Horndeski:1974wa}
\begin{equation}
\mathcal{L}=\frac{1}{16\pi}\Big( (1+{\beta}\sqrt{-X})R-2\Lambda +\eta X -\frac{\beta}{2\sqrt{-X}} [(\Box\phi)^2-(\nabla_\mu\nabla_\nu\phi)^2]\Big),
\end{equation}
where $\beta$ and $\eta$ are constants. 

\subsubsection{Solution}
A black hole solution for this theory is introduced in Ref.  \cite{Babichev:2017guv} with the metric
\begin{equation}
\mrd s^2=-f(r)\mrd t^2+\frac{\mrd r^2}{f(r)}+r^2(\mrd\theta^2+\sin^2\theta \mrd\varphi^2),
\end{equation}
and
\begin{align}
f=1-\frac{2m}{r}-\frac{\beta^2}{2\eta r^2}-\frac{\Lambda r^2}{3}, \quad \mrd\phi=\frac{\sqrt{2}\beta}{\eta r^2 \sqrt{f}} \mrd r.
\end{align}
The cosmological gauge field for this solution is (see Appendix \ref{app A})
\begin{equation}\label{A Horndeski 4dim}
A=-\frac{\sqrt{|\Lambda|}}{3}r^3\sin \theta\, \mrd t \wedge\mrd \theta \wedge\mrd \varphi,    
\end{equation}
which is fixed in a gauge such that it reproduces the mass correctly using the covariant formulation of charges in Appendix \ref{appendix cov}.

\subsubsection{Properties} 
The mass, temperature and entropy for this solution are \cite{Hajian:2020dcq}
\begin{equation}
M= m, \qquad T_{_{\text{H}}}=\frac{\beta^2+2\eta (r_{_\text{H}}^2-\Lambda r^4_{_\text{H}})}{8\pi\eta r^3_{_\text{H}}}, \qquad S=\pi r_{_\text{H}}^2,
\end{equation}
respectively, and the cosmological charge and potential are, respectively,
\begin{equation} \label{Theta 4.2}
C=-\frac{\sqrt{|\Lambda|}}{4\pi}, \qquad \Theta_\H=-\frac{\sqrt{|\Lambda|}4\pi r_\H^3}{3}.  
\end{equation}
\subsubsection{The first law and the Smarr formula} 
This solution has two free parameters $m$ and $\Lambda$. The generalized first law for this solution is
\begin{align}
&\delta M=T_\H \delta S_\H+\Theta_\H \delta C,
\end{align}
which can be checked to be a correct relation by using variations with respect to the two free parameters of this solution. For this solution, the generalized Smarr formula is \emph{not} satisfied as in some of the examples above.

\subsection{A black brane in Horndeski gravity}
\subsubsection{Theory}
The Lagrangian of the theory is \cite{Horndeski:1974wa}
\begin{equation}\label{Horndeski L 2}
\mathcal{L}=\frac{1}{16\pi}\Big(R-2\Lambda-F_{\mu\nu}F^{\mu\nu}-2(\alpha g_{\mu\nu}-\gamma G_{\mu\nu})\nabla^\mu\phi\nabla^\nu\phi\Big)
\end{equation}
in which  $G_{\mu\nu}=R_{\mu\nu}-\frac{1}{2}Rg_{\mu\nu}$ is the Einstein tensor. 

\subsubsection{Solution}
The metric in the coordinates $x^\mu=(t,r,x,y)$ is
\begin{align}
&\mrd s^2=-h(r)\mrd t^2+\frac{\mrd r^2}{f(r)}+r^2(\mrd x^2+ \mrd y^2),\\
&h=\frac{r^2}{\ell^2}-\frac{m}{r}+\frac{4q^2}{(4+\beta)r^2}-\frac{4q^4\ell^2}{15(4+\beta)^2r^6}, \nonumber\\
&f=\frac{(4+\beta)^2r^8 h}{\big(\frac{2q^2\ell^2}{3}-(4+\beta)r^4\big)^2}, \nonumber\\
& \mrd\phi=\sqrt{\frac{\beta-\frac{2q^2\ell^2}{3r^4}}{4\gamma f}}\mrd r, \quad A=\left(\frac{q}{r}-\frac{2q^3\ell^2}{15(4+\beta) r^5} \right) \mrd t,
\end{align}
with \cite{Feng:2015wvb}
\begin{equation}
\Lambda=-\frac{3(1+\frac{\beta}{2})}{\ell^2}, \qquad \alpha=\frac{3\gamma}{\ell^2}.
\end{equation}
It is easy to see that in order to vary $\Lambda$ while keeping the $\alpha$ fixed, one can simply take variations with respect to $\beta$. So, in order to check the first law, we will use variations with respect to $\beta$ which appears explicitly in the solution, instead of the $\Lambda$. The cosmological gauge field for this solution is (see Appendix \ref{app A})
\begin{equation}\label{A Horndeski brane}
\boldsymbol{A}=-\sqrt{|\Lambda|}\left(\frac{r^3}{3}+\frac{2q^2\ell^2}{3r(4+\beta)}-\frac{m\ell^2}{6}\right)\,\mrd t \wedge\mrd \theta \wedge\mrd \varphi.    
\end{equation}
The first two terms in the parentheses are determined by the equation $F=\mrd \boldsymbol A$ and Eq.\eqref{F on-shell}, while the last term in the parentheses is a gauge-fixing term; i.e., it does not contribute to $F$ by the equation $F=\mrd \boldsymbol A$. This gauge-fixing term is determined by putting the $\boldsymbol A$ and its variations into the covariant formulation of charges to reproduce mass correctly.

\subsubsection{Properties} 
The mass, electric charge, and entropy ``densities" for this solution are \cite{Hajian:2020dcq}, respectively,
\begin{equation}
M=\frac{(4+\beta)m}{32\pi}, \qquad Q=\frac{q}{4\pi}, \qquad S=\frac{r_\H^2}{4}. 
\end{equation}
By densities, it is understood that the charges are calculated without performing the integration over the $x$ and $y$ coordinates. Besides, the surface gravity and electric potential on the horizon are, respectively, 
\begin{equation}
 \kappa=\frac{3r_{_\text{H}}}{2\ell^2}-\frac{q^2}{(4+\beta)r^3_{_\text{H}}}  , \qquad \Phi_{_\text{H}}=\frac{q}{r_{_\text{H}}}-\frac{2q^3\ell^2}{15(4+\beta) r^5_{_\text{H}}}.   
\end{equation}
This example is a very special example in this work, because the standard (as well as the generalized) first law and Smarr formula do not hold if one uses the Hawking temperature $T_0=\frac{\kappa}{2\pi}$ as the temperature of the black brane. However, in Ref. \cite{Hajian:2020dcq}, it is shown that this is a generic feature in Horndeski gravity (and any model of gravity in which the speed of graviton differs from $c=1$). The physical temperature in Hawking radiation is dominated by the gravitons, and it is related to the Hawking temperature by an overall factor which is a function of the parameters of the solution. The interested reader is invited to study the original paper \cite{Hajian:2020dcq} for the details. Here, we report only the final result for the example under consideration. The physical temperature is related to the $T_0$ by 
\begin{equation}
T_\H= \left(\frac{3(4 +\beta)r_\H ^4-2q^2\ell^2 }{12 r^4_{_\text{H}}}\right) T_0.
\end{equation}
The cosmological charge and potential for this solution are, respectively,
\begin{equation} \label{Theta 4.3}
C=-\frac{\sqrt{|\Lambda|}}{4\pi}, \qquad \Theta_\H=-\sqrt{|\Lambda|} \left(\frac{r_\H^3}{3}+\frac{2q^2\ell^2}{3r_\H(4+\beta)}-\frac{m\ell^2}{6}\right).  
\end{equation}

\subsubsection{The first law and the Smarr formula} 
This solution has three free parameters $m$, $q$, and $\beta$. This latter parameter is representative of the $\Lambda$ in the solution. The generalized first law for this solution is
\begin{align}
&\delta M=T_\H \delta S_\H+\Phi_\H\delta Q+\Theta_\H \delta C,
\end{align}
which can be checked to be a correct relation by using variations with respect to the three free parameters of this solution. For this solution, the generalized Smarr formula is \emph{not} satisfied as like some of the previous examples.   

\subsection{Martinez-Teitelboim-Zanelli (MTZ) black hole}
\subsubsection{Theory}
The Lagrangian has the metric $g_{\mu\nu}$, a scalar field $\phi$, and the Maxwell gauge field $A_\mu$ as dynamical fields \cite{Martinez:2002ru,Barlow:2005yd}:
\begin{equation}
\mathcal{L}=\frac{1}{16\pi }\left(R-2\Lambda-F_{\mu\nu}F^{\mu\nu}-2\nabla_\mu \phi \nabla^\mu \phi -\frac{1}{3}R\phi^2-\alpha \phi^4\right).
\end{equation}
\subsubsection{Solution}
The dynamical fields in the coordinates $x^\mu=(t,r,\theta,\varphi)$ are \cite{Martinez:2002ru,Barlow:2005yd} 
\begin{align}\label{MTZ}
&\mrd s^2= -f\mrd t^2 +\frac{\mrd r^2}{f}+r^2 (\mrd \theta^2 + \sin^2\theta \mrd \varphi^2), \qquad f=(1-\frac{m}{r})^2-\frac{r^2}{\ell^2},\nonumber\\
&A=\frac{q}{r}, \qquad \phi=\frac{\sqrt{3(m^2-q^2)}}{r-m},
\end{align}
where
\begin{equation}\label{MTZ q relation}
\Lambda=\frac{3}{\ell^2}, \, \qquad q^2=m^2(1+\frac{2\Lambda}{9\alpha}).    
\end{equation}
Horizon radii are at  $r_\pm=\frac{\ell}{2}(\pm 1\mp\sqrt{1\mp\frac{4m}{\ell}})$, and the cosmological horizon is at $r_c=\frac{\ell}{2}(1+\sqrt{1-\frac{4m}{\ell}})$. It is clear that in order to have black holes, the conditions $0<m<\frac{\ell}{4}$ and $\alpha<\frac{-2\Lambda}{9}$ should be satisfied. Moreover, $0<\Lambda$ should have de Sitter asymptotics for this solution. In our analysis, we will focus on  $r_\H=r_+$, i.e. the black hole event horizon. However, the analysis applies to the other horizons by inserting an appropriate sign for the temperature. The cosmological gauge field $\boldsymbol A$ in a gauge which is fixed similar to the other examples mentioned above, can be found to be (see Appendix \ref{app A})
\begin{equation}
\boldsymbol{A}=-\frac{\sqrt{\Lambda} r^3}{3}\,\mrd t \wedge\mrd \theta \wedge\mrd \varphi.    
\end{equation}

\subsubsection{Properties} 
The mass, electric charge and horizon potential, temperature, and the entropy of MTZ black hole can be found,  respectively, as \cite{Martinez:2002ru,Winstanley:2004ay,Barlow:2005yd}
\begin{align}
&M=m,\qquad Q=q, \qquad \Phi_\H=\frac{q}{r_\H}, \qquad   T_\H=\frac{m(r_\H-m)}{2\pi r_\H^3}-\frac{\Lambda r_\H}{6\pi}, \qquad S_\H=\pi r_\H^2\left(1-\frac{m^2-q^2}{(r_\H-m)^2}\right).
\end{align}
We notice that the temperature is the standard Hawking temperature, which can be found by the relation $T_\H=\frac{1}{4\pi}\frac{df}{dr}$ on the horizon, while the entropy is the Bekenstein-Hawking entropy $\frac{A_\H}{4}$ multiplied by the factor of scalar curvature $R$ in the Lagrangian, i.e., $1-\frac{\phi^2}{3}$. The cosmological charge and potential are, respectively, 
 \begin{equation} \label{Theta 4.4}
C=\frac{\sqrt{\Lambda}}{4\pi}, \qquad \Theta_\H= -\frac{\sqrt{\Lambda} 4\pi r_H^3}{3}.  
\end{equation}   

\subsubsection{The first law and the Smarr formula} 
This solution has three parameters $m$, $q$, and $\ell$, but $q$ is not an independent parameter and is related to the other two parameters by the relation \eqref{MTZ q relation}.  The generalized first law for this solution is
\begin{align}
&\delta M=T_\H \delta S_\H+\Phi_\H\delta Q+\Theta_\H \delta C,
\end{align}
which can be checked to be a correct relation by using variations with respect to the two free parameters of this solution. For this solution, the generalized Smarr formula is \emph{not} satisfied.

\section{Solutions in five and higher dimensions}

\subsection{(A)dS-Myers-Perry black hole}
The (A)dS-Myers-Perry black hole solution is a generalization of the (A)dS-Kerr black hole  to five (and higher) dimensions \cite{Myers:1986un}. 

\subsubsection{Theory}
Einstein-$\Lambda$ gravity in $5$ dimension 
\begin{equation}
\mathcal{L}=\frac{1}{16\pi }(R-2\Lambda).
\end{equation}

\subsubsection{Solution}
The metric in the coordinates $x^\mu=(t,r,\theta,\varphi,\psi)$ with $\theta\in [0,\frac{\pi}{2}]$ and $\varphi , \psi \in [0,2\pi]$ is
\begin{align}
\mathrm{d}s^2= -& \frac{\Delta_\theta(1-\frac{\Lambda r^2}{6}) \mrd t^2}{\Xi_a\Xi_b}+\frac{2m}{\rho^2}(\frac{\Delta_\theta \mrd t}{\Xi_a\Xi_b}-a^2\sin^2\theta\frac{\mrd \varphi}{\Xi_a}-b^2\cos^2\theta \frac{\mrd \psi}{\Xi_b})^2\nonumber\\
&+\frac{\rho^2 \mrd r^2}{\Delta_r}+\frac{\rho^2\mrd \theta^2}{\Delta_\theta}
+\frac{r^2+a^2}{\Xi_a}\sin^2\theta \mrd \varphi^2 +\frac{r^2+b^2}{\Xi_b}\cos^2\theta \mrd \psi^2,
\end{align}
where
\begin{align}
\Delta_r&=\frac{(r^2+a^2)(r^2+b^2)(1-\frac{\Lambda r^2}{6})}{r^2}-2m, \qquad \Delta_\theta=1+\frac{a^2\Lambda}{6}\cos^2\theta+\frac{b^2\Lambda}{6}\sin^2\theta, \nonumber \\
\rho^2&= r^2+a^2 \cos^2\theta +b^2 \sin^2\theta,\qquad \Xi_a=1+\frac{a^2\Lambda}{6}, \qquad \Xi_b=1+\frac{b^2\Lambda}{6}.
\end{align}
Horizons of the Myers-Perry black hole are situated at $r_\H$ which are the roots of $\Delta_r=0$. The cosmological gauge field can be found to be (see Appendix \ref{app A})
\begin{align}
&A=-\frac{\sqrt{|\Lambda|}\sin\theta\cos\theta}{\Xi_a\Xi_b}\left(\frac{r^4+2r^2(a^2\cos^2\theta+b^2\sin^2\theta)}{4}+\alpha_0\right) \,\mrd t \wedge\mrd \theta \wedge\mrd \varphi \wedge \mrd\psi, \\
&\alpha_0=\frac{a^2b^2}{4}+\frac{m(a^2+b^2+\frac{a^2b^2\Lambda}{3})}{6\Xi_a\Xi_b}. \nonumber
\end{align}
The constant $\alpha_0$, which is a gauge-fixing term, is determined by the covariant formulation of charges in Appendix \ref{appendix cov}.

\subsubsection{Properties}
Denoting the angular momenta associated with the axial symmetries of the coordinates $\varphi$ and $\psi$ by $J_\varphi$ and $J_\psi$,
\begin{align}\label{Prop 5.1}
&M=\frac{\pi m (2\Xi_a+2\Xi_b-\Xi_a\Xi_b)}{4\Xi_a^2\Xi_b^2}, \qquad J_\varphi = \frac{\pi am}{2\Xi_a^2\Xi_b}, \qquad J_\psi = \frac{\pi bm}{2\Xi_a\Xi_b^2}, \nonumber\\
&\Omega_{\H}^{\varphi}=\frac{a(1-\frac{\Lambda r_\H^2}{6})}{(r_\H^2+a^2)}, \qquad \Omega_{\H}^{\psi}=\frac{b(1-\frac{\Lambda r_\H^2}{6})}{(r_\H^2+b^2)},\nonumber\\
& T_\H=\frac{r_\H^4[1-\frac{\Lambda}{6}(2r_\H^2+a^2+b^2)]-a^2b^2}{2\pi r_\H[(r_\H^2+a^2)(r_\H^2+b^2)]}, \qquad    S_\H= \frac{\pi^2[(r_\H^2+a^2)(r_\H^2+b^2)]}{2\Xi_a\Xi_b r_\H}.
\end{align}
The cosmological charge and potential can be read from Eqs. \eqref{C} and \eqref{Theta H}:
 \begin{equation} \label{Theta 5.1}
C=\pm\frac{\sqrt{|\Lambda|}}{4\pi}, \qquad \Theta_\H= -\frac{\sqrt{|\Lambda|}\pi^2}{\Xi_a\Xi_b}\left(\frac{(r_\H^2+a^2)(r_\H^2+b^2)}{2}+\frac{m(a^2+b^2+\frac{\Lambda a^2 b^2}{3})}{3\Xi_a\Xi_b}\right),  
\end{equation}   
with the positive and negative $C$ for the solutions with dS and AdS asymptotics. 

\subsubsection{The first law and the Smarr formula} 
This solution has four free parameters $(m,a,b,\Lambda)$.  Using Eqs. \eqref{Prop 5.1} and \eqref{Theta 5.1}, the generalized first law and Smarr formula for this solution 
\begin{align}
&\delta M=T_\H \delta S_\H+\Omega^\varphi_\H\delta J_\varphi+\Omega_\H^\psi\delta J_\psi+\Theta_\H \delta C, \label{First law 5.1}\\
&2M = 3T_\H S_\H +3\Omega^\varphi_\H J_\varphi+3\Omega_\H^\psi J_\psi- \Theta_\H C, \label{Smarr 5.1}
\end{align}
respectively, can be checked for variations with respect to the parameters $p_i\in \{m,a,b,\Lambda\}$. Hence, for this example the generalized first law and Smarr formula hold. For the solutions  whose horizon may not be found analytically in terms of the parameters of the solution (like Myers-Perry solutions), we refer the reader to Appendix \ref{app first law Smarr}, in order to find how to check the first law and Smarr formula easily. 

\subsection{(A)dS-Reissner-Nordstr$\ddot{\text{o}}$m-Tangherlini black hole} \label{Tanghrlini sec}
This family of black holes is a generalization of the (A)dS-Reissner-Nordstr$\ddot{\text{o}}$m black hole to higher $D$ dimensions, which are spherically symmetric solutions with electric charges. 

\subsubsection{Theory}
With the dynamical fields as the metric $g_{\mu\nu}$ and Maxwell gauge field $A_\mu$, the theory is described by the Lagrangian of Einstein-Maxwell-$\Lambda$ gravity in $D$ dimensions, and the metric reads  
\begin{equation}
\mathcal{L}=\frac{1}{16\pi }(R-2\Lambda-F_{\mu\nu}F^{\mu\nu}),
\end{equation}
where $F_{\mu\nu}=\partial_\mu A_\nu-\partial_\nu A_\mu$ is the field strength.

\subsubsection{Solution}
Denoting the time and radius coordinates by $t$ and $r$, respectively, for these black holes in $D$ dimensions, 
\begin{align}
& \mathrm{d}s^2  =-f\mrd t^2+\frac{\mrd r^2}{f}  +r^2
\mrd \Omega_{_{D-2}}^2, \qquad A=\sqrt{\frac{(D-2)}{2(D-3)}}\frac{q}{r^{D-3}}\mrd t,\nonumber\\
 & f=1-\frac{2m}{r^{D-3}}+\frac{q^2}{r^{2(D-3)}}-\frac{2\Lambda r^2}{(D-1)(D-2)}, \qquad \Omega_{_{D-2}}=\frac{2\pi^{\frac{D-1}{2}}}{\Gamma(\frac{D-1}{2})}\,, 
\end{align}
where $\Omega_{_{D-2}}$ is the area of the $D-2$-dimensional unit sphere and $\Gamma$ is the gamma function. Horizons are situated at the radii which can be found as roots of $f(r_\H)=0$.
The cosmological gauge field for this family of solutions is (see Appendix \ref{app A})
\begin{equation}\label{A RN Tangh}
\boldsymbol A=-\frac{\sqrt{|\Lambda|}}{D-1}r^{D-1} \, \mrd t \wedge \mrd \Omega_{_{D-2}}.    
\end{equation}

\subsubsection{Properties}
For these black holes, mass, electric charge and potential, temperature and entropy are, respectively, 
\begin{align}\label{Prop 5.2}
&M=\frac{(D-2)\Omega_{_{D-2}} m}{8 \pi}, \qquad Q=\frac{\sqrt{\frac{(D-2)(D-3)}{2}}\Omega_{_{D-2}} q}{4\pi},\qquad \Phi_\H= \sqrt{\frac{D-2}{2(D-3)}}\frac{q}{r_\H^{D-3}},\nonumber \\
& T_\H=\frac{1}{4\pi}\left(\frac{2(D-3)m}{r_\H^{D-2}}-\frac{2(D-3)q^2}{r_\H^{2(D-3)+1}}-\frac{4\Lambda r_\H}{(D-1)(D-2)}\right), \qquad    S_\H= \frac{r_\H^{D-2}\Omega_{_{D-2}}}{4}.
\end{align}
The temperature is the standard Hawking temperature which can be found by the relation $T_\H=\frac{1}{4\pi}\frac{df}{dr}$ on the horizon, while the entropy is the Bekenstein-Hawking entropy $\frac{A_\H}{4}$. Using Eqs. \eqref{C} and \eqref{Theta H}, the cosmological charge and potential are found as
 \begin{equation} \label{Theta 5.2}
C=\pm\frac{\sqrt{|\Lambda|}}{4\pi}, \qquad \Theta_\H= -\frac{\sqrt{|\Lambda|}}{D-1}{r_\H^{D-1} \Omega_{_{D-2}}},  
\end{equation}   
with the positive and negative $C$ for the solutions with dS and AdS asymptotics.

\subsubsection{The first law and the Smarr formula}
The Reissner-Nordstr$\ddot{\text{o}}$m-Tangherlini black holes have three free parameters $(m,q,\Lambda)$.  Using Eqs.  \eqref{Prop 5.2} and \eqref{Theta 5.2} and variations with respect to the three parameters, the generalized first law and Smarr formula for this family of solutions are satisfied, respectively,  as 
\begin{align}
&\delta M=T_\H \delta S_\H+\Phi_\H \delta Q+\Theta_\H \delta C, \label{First law 5.2}\\
&(D-3)M = (D-2)T_\H S_\H +(D-3)\Phi_\H Q- \Theta_\H C. \label{Smarr 5.2}
\end{align}
For these solutions the horizon radii may not be found analytically in terms of the parameters of the solution. We refer the reader to the Appendix \ref{app first law Smarr}, in order to find how to check the first law and Smarr formula without having the explicit form of $r_\H$.

\subsection{Charged rotating black hole in minimal gauged supergravity}
\subsubsection{Theory}
The Lagrangian of the minimal gauged supergravity in five dimensions is 
\begin{equation}
\mathcal{L}=\frac{1}{16\pi }(R-2\Lambda-F_{\mu\nu}F^{\mu\nu}+\frac{2}{3\sqrt{3}}\epsilon^{\mu_1\mu_2\dots \mu_5}F_{\mu_1\mu_2}F_{\mu_3\mu_4}A_{\mu_5}),    
\end{equation}
where $\epsilon_{\mu_1\mu_2\dots \mu_5}$ is the five-dimensional Levi-Civita symbol with components $+1$ or $-1$. The last term in the Lagrangian above is the Chern-Simons term.

\subsubsection{Solution}
The metric in the coordinates $x^\mu=(t,r,\theta,\varphi,\psi)$ with $\theta\in [0,\frac{\pi}{2}]$ and $\varphi , \psi \in [0,2\pi]$ is \cite{Chong:2005hr}
\begin{align}
\mathrm{d}s^2= -& \frac{\Delta_\theta[(1-\frac{\Lambda r^2}{6})\rho^2 \mrd t+2q\nu]\mrd t}{\Xi_a\Xi_b\rho^2}+\frac{2q\nu\omega}{\rho^2}+\frac{f}{\rho^4}(\frac{\Delta_\theta \mrd t}{\Xi_a\Xi_b}-\omega)^2+\frac{\rho^2 \mrd r^2}{\Delta_r}+\frac{\rho^2\mrd \theta^2}{\Delta_\theta}\nonumber\\
&+\frac{r^2+a^2}{\Xi_a}\sin^2\theta \mrd \varphi^2 +\frac{r^2+b^2}{\Xi_b}\cos^2\theta \mrd \psi^2,
\end{align}
where
\begin{align}
\nu & =b\sin^2\theta\mrd \varphi+a\cos^2\theta \mrd \psi, \qquad \omega=a\sin^2\theta\frac{\mrd\varphi}{\Xi_a}+b\cos^2\theta\frac{\mrd\psi}{\Xi_b}, \qquad f=2m\rho^2-q^2-\frac{\Lambda}{3}abq\rho^2\nonumber \\
\Delta_r&=\frac{(r^2+a^2)(r^2+b^2)(1-\frac{\Lambda r^2}{6})+q^2+2abq}{r^2}-2m, \qquad \Delta_\theta=1+\frac{a^2\Lambda}{6}\cos^2\theta+\frac{b^2\Lambda}{6}\sin^2\theta, \nonumber \\
\rho^2&= r^2+a^2 \cos^2\theta +b^2 \sin^2\theta,\qquad \Xi_a=1+\frac{a^2\Lambda}{6}, \qquad \Xi_b=1+\frac{b^2\Lambda}{6}.
\end{align}
The Maxwell gauge field in this solution is 
\begin{equation}
A=\frac{\sqrt{3}q}{2\rho^2}(\frac{\Delta_\theta \mrd t}{\Xi_a\Xi_b}-\omega).    
\end{equation}
In the special cases of $q=0$ and $a=b=0$, one recovers the (A)dS-Myers-Perry and the  (A)dS-Reissner-Nordstr$\ddot{\text{o}}$m-Tangherlini black holes in five dimensions. However, in its general form, it is not a solution to Einstein-Maxwell-$\Lambda$ theory. Instead, it is a solution to a theory which is supersymmetric and has a Chern-Simons term in it. Having in mind that conserved charges depend on the Lagrangian, it is worth studying this solution separately. The horizon radii are 
\begin{equation}
r_\pm^2=m-\frac{a^2+b^2}{2}\pm\sqrt{(m-\frac{(a-b)^2}{2}+q)(m-\frac{(a+b)^2}{2}-q)}.    
\end{equation}
The cosmological gauge field can be found to be (see appendix \ref{app A})
\begin{align}
&\boldsymbol A=-\frac{\sqrt{|\Lambda|}\sin\theta\cos\theta}{\Xi_a\Xi_b}\left(\frac{r^4+2r^2(a^2\cos^2\theta+b^2\sin^2\theta)}{4}+\sigma_0\right) \,\mrd t \wedge\mrd \theta \wedge\mrd \varphi \wedge \mrd\psi, \\
&\sigma_0=\frac{a^2b^2}{4}+\frac{m(a^2+b^2+\frac{a^2b^2\Lambda}{3})}{6\Xi_a\Xi_b}+\frac{abq(\Xi_a+\Xi_b)}{3\Xi_a\Xi_b}. \nonumber
\end{align}
The constant $\alpha_0$, which is a gauge fixing term, is determined by the covariant formulation of charges in the Appendix \ref{appendix cov}.

\subsubsection{Properties}
The thermodynamic properites of this solution are\cite{Chong:2005hr}
\begin{align}\label{Prop 5.3}
&M=\frac{\pi m (2\Xi_a+2\Xi_b-\Xi_a\Xi_b)-2\pi abq\frac{\Lambda}{6}(\Xi_a+\Xi_b)}{4\Xi_a^2\Xi_b^2}, \qquad Q=\frac{\sqrt{3}\pi q}{2\Xi_a\Xi_b},\nonumber\\
& J_\varphi = \frac{\pi(2am+qb(1-\frac{a^2\Lambda}{6}))}{4\Xi_a^2\Xi_b}, \qquad J_\psi = \frac{\pi(2bm+qa(1-\frac{b^2\Lambda}{6}))}{4\Xi_a\Xi_b^2}, \nonumber\\
&\Omega_{\H}^{\varphi}=\frac{a(r_\H^2+b^2)(1-\frac{\Lambda r_\H^2}{6})+bq}{\sigma}, \qquad \Omega_{\H}^{\psi}=\frac{b(r_\H^2+a^2)(1-\frac{\Lambda r_\H^2}{6})+aq}{\sigma}, \qquad \Phi_\H= \frac{\sqrt{3} q r_\H^2}{2\sigma},\nonumber\\
& T_\H=\frac{r_\H^4[1-\frac{\Lambda}{6}(2r_\H^2+a^2+b^2)]-(ab+q)^2}{2\pi r_\H\sigma}, \qquad    S_\H= \frac{\pi^2\sigma}{2\Xi_a\Xi_b r_\H},
\end{align}
and $\sigma = (r_\H^2+a^2)(r_\H^2+b^2)+abq$.
The cosmological charge and potential by Eqs. \eqref{C} and \eqref{Theta H} are
 \begin{equation} \label{Theta 5.3}
C=\pm\frac{\sqrt{|\Lambda|}}{4\pi}, \qquad \Theta_\H= -\frac{\sqrt{|\Lambda|}\pi^2}{\Xi_a\Xi_b}\left(\frac{(r_\H^2+a^2)(r_\H^2+b^2)}{2}+\frac{m(a^2+b^2+\frac{\Lambda a^2 b^2}{3})+2abq(\Xi_a+\Xi_b)}{3\Xi_a\Xi_b}\right),  
\end{equation}   
respectively, with the positive and negative $C$ for the solutions with dS and AdS asymptotics. 

\subsubsection{The first law and the Smarr formula}
This solution has five free parameters $(m,a,b,q,\Lambda)$.  Using Eqs. \eqref{Prop 5.3} and \eqref{Theta 5.3}, the generalized first law and Smarr formula for this solution 
\begin{align}
&\delta M=T_\H \delta S_\H+\Omega^\varphi_\H\delta J_\varphi+\Omega_\H^\psi\delta J_\psi+\Phi_\H\delta Q+\Theta_\H \delta C, \label{First law 5.3}\\
&2M = 3T_\H S_\H +3\Omega^\varphi_\H J_\varphi+3\Omega_\H^\psi J_\psi + 2\Phi_\H Q - \Theta_\H C, \label{Smarr 5.3}
\end{align}
respectively, can be checked for variations with respect to the parameters $p_i\in \{m,a,b,q,\Lambda\}$. Hence, for this example the generalized first law and Smarr formula hold.

\subsection{Lifshitz $z=2$ black brane}
\subsubsection{Theory}
The Lagrangian contains second-order terms in curvature as follows:
\begin{equation}
\mathcal{L}=\frac{1}{16\pi }\left(R-2\Lambda+\alpha R^2 + \beta R_{\mu\nu}R^{\mu\nu} + \gamma ( R^2 -4 R_{\mu\nu}R^{\mu\nu}+R_{\mu\nu\sigma\rho}R^{\mu\nu\sigma\rho})\right).
\end{equation}
The last term is the Gauss-Bonnet term, and the coupling constants are
\begin{equation}
\Lambda=-\frac{2197}{551 \ell^2}, \quad \alpha = -\frac{16\ell^2}{725}, \quad   \beta=\frac{1584 \ell^2}{13775}, \quad \gamma=\frac{2211 \ell^2}{11020}.  
\end{equation}
\subsubsection{Solution}
The metric in the coordinates $x^\mu=(t,r, x, y ,z)$ is \cite{AyonBeato:2010tm, Gim:2014nba} 
\begin{equation}
\mrd s^2=-(\frac{r}{\ell})^{2z}(1-\frac{m\ell^{\frac{5}{2}}}{r^{\frac{5}{2}}})\,\mrd t^2 +\frac{\mrd r^2}{\frac{r^2}{\ell^2}(1-\frac{m\ell^{\frac{5}{2}}}{r^{\frac{5}{2}}})}+r^2 (\mrd x^2+\mrd y^2+\mrd z^2)
 \end{equation} 
for $z=2$. The horizon is a brane located at $r_\H=m^{\frac{2}{5}}\ell$. The cosmological gauge field for this solution is (see Appendix \ref{app A})
\begin{equation}
A = - \sqrt{|\Lambda|} (\frac{r^5}{5\ell}-\frac{13121 m^2\ell^4}{87880} )\, \mrd t \wedge \mrd x \wedge \mrd y \wedge \mrd z.
\end{equation}
The last term is a gauge-fixing term which is fixed by using the covariant charge method of charges (see Appendix \ref{appendix cov}) to reproduce the $\delta_{\ell}M$ correctly.

\subsubsection{Properties}
Using the solution phase space method in Refs. \cite{HS:2015xlp,Ghodrati:2016vvf} or other methods \cite{Hohm:2010jc,Gonzalez:2011nz,Gim:2014nba,Ayon-Beato:2015jga} we find
\begin{align}
&M=\frac{297 m^2 \ell^2}{17632\pi},\qquad \quad T_\H=\frac{5 m^{\frac{4}{5}}}{8\pi\ell}, \qquad \quad {S}_\H=\frac{99m^{\frac{6}{5}}\ell^3}{2204}.
\end{align}
$M$ and ${S}_\H$ denote mass and entropy densities, respectively,  of the black brane. 
Using Eqs.  \eqref{C} and \eqref{Theta H}, the cosmological charge and potential are:
\begin{equation}
C=-\frac{\sqrt{-\Lambda}}{4\pi}, \qquad    \Theta_\H = - \sqrt{|\Lambda|} (\frac{r_\H^5}{5\ell}-\frac{13121 m^2\ell^4}{87880}).
\end{equation}
\subsubsection{The first law and the Smarr formula}
This solution has two free parameters $m$ and $\ell$. The generalized first law and Smarr formula for this solution are
\begin{align}
&\delta M=T_\H \delta S_\H+\Theta_\H \delta C,\\
&2M=3T_\H S_\H-\Theta_\H C,
\end{align}
which can be checked to be a correct relation by using variations with respect to the two free parameters of this solution. We note that the couplings $(\alpha, \beta,\gamma)$ are not independent from the $\Lambda$. So, we could expect to have the Smarr formula without contributions from  these parameters.

\subsection{AdS-Schwarzschild black holes in higher curvature gravity}
\subsubsection{Theory}
The Lagrangian which we consider as the last example in this work is the Einstein-$\Lambda$ gravity with higher curvature terms in arbitrary $D>2$ dimensions
\begin{equation}
\mathcal{L}=\frac{1}{16\pi}\Big(R-2\Lambda+\alpha R^2+\beta R_{\mu\nu}R^{\mu\nu}\Big),
\end{equation}
in which $\alpha$ and $\beta$ are arbitrary constants.

\subsubsection{Solution}
The metric is simply a generalization of AdS-Schwarzschild black hole to $D$ dimensions, which is
\begin{equation}
\mrd s^2=-f\mrd t^2+\frac{\mrd r^2}{f}+r^2 \mrd \Omega_{_{D-2}}^2, \qquad f=1-\frac{2m}{r^{D-3}}+\frac{r^2}{\ell^2},
\end{equation}
where $\ell$ satisfies $\Lambda =\frac{-\ell^2(D^2-3D+2)+(\alpha D + \beta)(D-4)(D-1)^2}{2\ell^4}$.
The cosmological gauge field for this family of solutions is (see Appendix \ref{app A})
\begin{equation}\label{A AdS-Schw-higher}
A=-{\sqrt{|\Lambda|}}\left(\frac{r^{D-1}}{D-1}+\sigma_0 \right) \, \mrd t \wedge \mrd \Omega_{_{D-2}}, \qquad \sigma_0= \frac{4m\ell^2(\alpha D +\beta)}{2(D-1)(D-4)(\alpha D+\beta)-(D-2)\ell^2}.  
\end{equation}
The $\sigma_0$ is a gauge fixing term which can be fixed by covariant formulation of charges which is described in Appendix \ref{appendix cov}.

\subsubsection{Properties}
Conserved charges, such as the mass and entropy, depend on the solution as well as the theory. As a result, although these black holes are simply the AdS-Schwarzschild solutions, but the theory differs from the Einstein-$\Lambda$ theory. The new charges associated with these solutions are different, and can be found to be \cite{Ghodrati:2016vvf}
\begin{align}
&M=\mathcal{X}\times \frac{(D-2)\Omega_{_{D-2}}}{8\pi} m,\qquad T_\H=\frac{(D-1)r_{_\text{H}}^{D-2}+(D-3)\ell^2\,r_{_\text{H}}^{D-4}}{4\pi\ell^2\,r_{_\text{H}}^{D-3}}, \qquad S_{_\text{H}}=\mathcal{X}\times \frac{r_{_\text{H}}^{D-2}\Omega_{_{D-2}}}{4}\nonumber
\end{align}
in which 
\begin{equation}\label{Prop. AdS-Schw-higher}
\mathcal{X}=\frac{\ell^2- 2D(D-1)\alpha- 2(D-1) \beta}{\ell^2}\,, \qquad \Omega_{_{D-2}}=\frac{2\pi^{\frac{D-1}{2}}}{\Gamma(\frac{D-1}{2})}\,,
\end{equation}
and horizons are determined by the equation $r_{_\text{H}}^{D-1}+\ell^2 r_{_\text{H}}^{D-3}-2m\ell^2=0$. By Eqs. \eqref{C} and \eqref{Theta H}, the cosmological charge and potential are, respectively,
\begin{equation}\label{Theta AdS-Schw-higher}
C=-\frac{\sqrt{-\Lambda}}{4\pi}, \qquad    \Theta_\H = - \sqrt{|\Lambda|} (\frac{r_\H^{D-1}}{D-1}+\sigma_0)\Omega_{_{D-2}}.
\end{equation}

\subsubsection{The first law and the Smarr formula}
This family of solutions has two free parameters in the solution $(m,\ell)$. These parameters should not be confused with the $(\alpha,\beta)$ which are free parameters or couplings in the Lagrangian. In the case of $\alpha=\beta=0$, we recover the AdS-Schwarzschild black holes in Einstein-$\Lambda$ theory which we have already studied in Sec. \ref{Tanghrlini sec} (by setting $q=0$). So, in this case, we have already shown that the generalized first law and Smarr formula hold. If at least one of the $\alpha$ or $\beta$ is nonzero, one can check that the first law is still satisfied, using Eqs.  \eqref{Prop. AdS-Schw-higher} and \eqref{Theta AdS-Schw-higher} and the method which is described in Appendix \ref{app first law Smarr} as
\begin{equation}
\delta M=T_\H \delta S_\H+\Theta_\H \delta C.    
\end{equation}
However, the Smarr formula fails to be satisfied, which is to be expected as one needs to incorporate the other dimensionful parameters $\alpha$ and/or $\beta$, which is an outstanding problem at this stage.

\section{Universality of the Smarr formula}
In the black hole physics literature, the Smarr formula is not considered as a universal relation. Clearly, it does depend on the dimension of spacetime $D$. However, one can still inquire if the Smarr relation \eqref{Smarr} is a generic relation. In spite of the fact that in some of the examples that we have analyzed this relation fails, one can see a suggestive pattern in it: This relation fails only for the Lagrangians which contain at least one free dimensionful parameter or coupling constant (in addition to the $\Lambda$). This observation suggests that this generalized Smarr formula should be extended such that it contains the contributions from those dimensionful parameters. In this regard, and based on our case-by-case study and the proof in Sec.  \ref{Smarr sec}, we put forward the following conjecture. 

\noindent\textbf{Conjecture:} \emph{The Smarr formula in Eq. \eqref{Smarr} can always be generalized to include contributions from dimensionful coupling constants in the Lagrangian. }

In order to do this generalization, one may  probably use a similar method as the one  used for  $\Lambda$. However, this is a subject of research beyond the scope of this paper and needs more investigations. Some guidelines for such an approach could be: (i) if the dimensionful parameter is a parameter in the Lagrangian, it should be promoted to be a parameter in the solution (not in the Lagrangian), probably as a conserved charge, and (ii) its conjugate chemical potential in the first law should be a (well-)defined property of the horizon; {i.e.,} it could be found using only the information in the vicinity of the horizon. 

Let us assume that such an analysis has been successfully done, yielding new conserved charges $K_i$ with dimensionality $K_i\sim l^{k^{(i)}}$ and their associated chemical potentials $\Psi^i_\H$, with the following contribution to the first law:
\begin{equation}\label{first law gen}
\delta M=T_{_\text{H}}\delta S+\Omega_{_\text{H}}\delta J+\Phi_{_\text{H}}\delta Q +\Theta_{_\text{H}}\delta C+\Psi^i_\H\delta K_i. \end{equation}
Following the steps in Sec. \ref{Smarr sec} verbatim, after scaling $l \to \alpha l$, one has
\begin{equation}\label{proof Smarr 2}
\alpha^{D-3}M(S,J,Q,C,K_i)=M\left(\alpha^{D-2}S,\alpha^{D-2}J,\alpha^{D-3}Q,\alpha^{-1}C,\alpha^{k^{(i)}}K_i\right).    
\end{equation}
Using the Euler relation \eqref{Euler} and Eq. \eqref{proof Smarr 2}, one gets
\begin{equation}
(D-3) M=(D-2)\left(\frac{\partial M}{\partial S}\right)S+(D-2)\left(\frac{\partial M}{\partial J}\right) J+ (D-3)\left(\frac{\partial M}{\partial Q}\right) Q -\left(\frac{\partial M}{\partial C}\right)C+k^{(i)}\left(\frac{\partial M}{\partial K^i}\right)K_i 
\end{equation}
in which the sum over $i$ is understood. At the end, using the generalized first law \eqref{first law gen}, we find the generalized Smarr relation 
\begin{equation}\label{Smarr gen}
{(D-3) M=(D-2)T_{_\text{H}}S+(D-2)\Omega_{_\text{H}} J+ (D-3)\Phi_{_\text{H}} Q -\Theta_{_\text{H}} C }+k^{(i)}\Psi^i_\H K_i.
\end{equation}

Having the general structure of the generalized Smarr formula, one may be interested to investigate and find $\Psi^i$ and $K_i$ for the examples which failed to satisfy the nongeneralized Smarr relation \eqref{Smarr}. This is a very interesting subject for research in the future, and is beyond the scope of this paper. Nonetheless, it is important to emphasize that, in order to find the correct contributions from dimensionful parameters to the Smarr relation, one needs to find a systematic and a precise description of these parameters as conserved charges (or at least as parameters of the solution); this is because 
\begin{itemize}
\item variation of a Lagrangian coupling constant in the first law is conceptually problematic.
\item The dimensional analysis may not determine $K_i$ uniquely. As an example, we remind the reader the difference of pressure $P$ in $V_{\text{eff}}\delta P$ compared to $C$ in $\Theta_\H \delta C$. The pressure (which is proportional to $\Lambda$) has dimension $l^{-2}$, while $C$ (which is proportional to $\sqrt{\Lambda}$) is of dimension $l^{-1}$. Nonetheless, both of $V_{\text{eff}}\delta P$ and $\Theta_\H \delta C$ are allowed by the dimensional analysis.
\item In the absence of a precise definition for the $\Psi^i_\H$, the first law (and consequently, the Smarr relation) could act only as a definition for it. Therefore, such relations would be trivially satisfied.
\end{itemize}
Accordingly, generalization of the first law and the Smarr relation for the problematic examples in this paper (without a systematic notion of charges and chemical potentials) can yield misleading outcomes, and thus we postpone their full study to later investigations.

\section{Conclusions}

The cosmological constant $\Lambda$ can be considered as a conserved charge $C$ associated with the gauge symmetry of a gauge field $A$. The conserved charge $C$ is analogous to electric charge: (i) It is a parameter of the solution, (ii) is extensive, and (iii) can be positive or negative. Besides, its conjugate $\Theta_\H$ is a property of the horizon. These properties resolve problems with the $V\delta P$ formulation of $\Lambda$ in the first law of black hole thermodynamics. In this paper, we generalized the Smarr formula to include a contribution from the $\Theta_\H C$ term, and provided a proof for it. However, the proof which is based on dimensional analysis, does not capture the free dimensionful parameters in the Lagrangian. We analyzed a handful number of examples to study this issue case by case. 

In addition, we showed that the $\Theta_\H$ reproduces the ``effective volume" successfully. Besides, we showed how the ambiguity of the effective volume can be removed by the role of gauge fixing in determination of $\Theta_\H$. Studying different examples in this paper collects a fair number of black holes with nonzero $\Lambda$ and can provide a reference for the readers about the cosmological gauge field $A$ as a part of the black hole solutions. 

The successful generalization of the first law for all of the examples, not only supports the $\Theta_\H\delta C$ formulation of $\Lambda$, but also it confirms the ``modified temperature" for Horndeski gravities which has been recently proposed in Ref. \cite{Hajian:2020dcq}.   

\noindent \textbf{Acknowledgements:} This work has been supported by  T$\ddot{\text{U}}$BITAK international researchers program No. 2221.

\appendix

\section{How to find the cosmological gauge field}\label{app A}
In this section, we present a heuristic method to find the cosmological gauge field. Let us denote the coordinates by $(t,r,x^1,\dots,x^{D-2})$ for the time, radius, and some other coordinates $x^i$. For black hole solutions which are stationary, components of the metric $g_{\mu\nu}$ can be chosen to be independent of $t$. So, the determinant of the metric $g$, could be a function of coordinates $(r,x^i)$. According to the Eq.  \eqref{F on-shell}, the cosmological gauge field strength is equal to 
\begin{equation}
F=\sqrt{|\Lambda|}\sqrt{-g}\, \mrd t \wedge \mrd r \wedge \mrd x^1 \wedge \cdots \wedge \mrd x^{D-2}.    
\end{equation}
The question is how to find a gauge field $A$ such that $F=\mrd A$. Up to a gauge transformation, the cosmological gauge field $A$ can be suggested to be
\begin{equation}
A=-\sqrt{|\Lambda|}\tilde g \,\, \mrd t \wedge  \mrd x^1 \wedge \cdots \wedge \mrd x^{D-2}, \qquad \tilde g=\int \mrd r \sqrt{-g}. 
\end{equation}
It can be easily checked that $F=\mrd A$ is satisfied. Besides, the constant of integration in $\tilde g$, which can be a function of parameters of the solutions as well as all coordinates except the $r$, is a part of the gauge freedom. This gauge freedom can be fixed by the covariant method of charges which is described in the next section. 

One could ask about other components for $A$, which are, in general, a linear combination of terms  $\mrd t \wedge \mrd r \wedge \mrd x^1 \wedge \cdots \wedge \mrd x^{D-2}$ with a missed $\mrd x^i$ and the term $\mrd r \wedge \mrd x^1 \wedge \cdots \wedge \mrd x^{D-2}$. The short answer is that such a component does not contribute to the $\Theta_\H$ defined in Eq. \eqref{Theta H}, because pullback of such a term  in the expression $\xi_\H\cdot A$ to the horizon vanishes, because such a term inevitably misses either a direction along $\mrd t$ to be contracted by $\xi_\H$ or one of $\mrd x^i$ to be integrated over the horizon.

\section{Covariant calculation of charges}\label{appendix cov}
In gravity theories, there are different methods for calculating conserved charges. Among the methods, one can mention some of the well-established methods like the Arnowitt-Deser-Misner (ADT) formulation \cite{Arnowitt:1959ah,Arnowitt:1960es,Arnowitt:1962hi} continued by Regge-Teitelboim \cite{Regge:1974zd}, Brown-York formulation \cite{Brown:1992br}, and ADT formulation of charges \cite{Abbott:1981ff, Deser:2002rt,Deser:2002jk}. In this paper, we have used a method which is called ``covariant formulation of charges" and has been introduced in the late 1980s and early 1990s \cite{Crnkovic:1987at, Ashtekar:1987hia,Ashtekar:1990gc, Lee:1990gr,Wald:1993nt,Iyer:1994ys,Wald:1999wa}. Interested reader can find reviews on this method in, e.g., Refs. \cite{Hajian:2015eha,Seraj:2016cym,sisman,Corichi:2016zac}. In this appendix, we briefly review the basics of this method and provide the final formula by which the charges are calculated. 

Phase space is a manifold with a 2-form, which is called symplectic form and is denoted by $\Omega$. The covariant phase space formulation of charges is based on a phase space which is built covariantly; instead of fields and their momentum conjugates in a time slice, the phase space is built by the fields over all of the spacetime which we denote collectively by $\Phi(x^\mu)$. So, we do not need to consider their momentum conjugates in the phase space. The symplectic 2-form of such a phase space is built as follows. Given a Lagrangian density $\mathcal{L}$, the surface term $\boldsymbol{\Theta}$ can be read by the variation of the Lagrangian dual $\mathbf{L}$:
\begin{equation}
\mrd \mathbf{L}=(\text{EOM})\delta \Phi+\mrd \boldsymbol{\Theta}(\delta \Phi,\Phi),    
\end{equation}
in which EOM denotes the equations of motions. Having the $\boldsymbol{\Theta}$ as a 1-form on the space of fields, and a $D-1$-form on space time, the symplectic current $\boldsymbol{\omega}$ is defined by 
\begin{equation}\label{omega}
\boldsymbol{\omega}(\delta_1\Phi,\delta_2\Phi,\Phi)=\delta_1\mathbf{\Theta}(\delta_2\Phi,\Phi)-\delta_2\mathbf{\Theta}(\delta_1\Phi,\Phi)\,,
\end{equation}   
which is just the exterior derivative of $\boldsymbol{\Theta}$ on the field configuration space. The symplectic 2-form which makes the field configuration space a phase space is 
\begin{equation}\label{Omega}
\Omega(\delta_1\Phi,\delta_2\Phi,\Phi)\equiv \int_\Sigma \boldsymbol{\omega}(\delta_1\Phi,\delta_2\Phi,\Phi)\, \end{equation}
where $\Sigma$ is a Cauchy surface. It can be shown that using appropriate boundary conditions, the result would not depend on the choice of this hypersurface. 

On the covariant phase space which is built by the procedure above, one can associate a charge variation $\delta H_\epsilon$ to a generator $\epsilon$. The generator can be a combination of diffeomorphisms and gauge transformations $\epsilon\equiv \{\xi^\mu, \lambda, \boldsymbol{\lambda}\}$. The diffeomorphism is $x^\mu\to x^\mu-\xi^\mu$, while $A\to A+\mrd\lambda$ and $\boldsymbol{A}\to\boldsymbol{A}+\mrd \boldsymbol{\lambda}$ are gauge transformations of the Maxwell field and cosmological gauge field respectively. Using the standard definition of charge variations in a phase space which is $\delta H_\epsilon\equiv \delta_\epsilon\Phi\cdot\Omega$, 
\begin{align}\label{delta H xi}
\delta H_{\epsilon}(\Phi)&\equiv  \int_\Sigma \big(\delta^{[\Phi]}\mathbf{\Theta}(\delta_\epsilon\Phi,\Phi)-\delta_\epsilon\mathbf{\Theta}(\delta\Phi,\Phi)\big)=\int_{\Sigma}\mrd\boldsymbol{k}_{\epsilon}(\delta\Phi,\Phi)=\oint_{\partial\Sigma}\boldsymbol{k}_{\epsilon}(\delta\Phi,\Phi) \,.
\end{align} 
In the equations above, the first equation is a result of $\mrd \boldsymbol\omega=0$ (onshell and for linearized perturbations), and the Poincar$\acute{\text{e}}$ lemma which admits $\boldsymbol\omega=\mrd \boldsymbol{k}$ for some $\boldsymbol{k}$. The last equation is the Stokes' theorem. The last equation is practically the most useful term for charge calculation in covariant formulation: For any solution $\Phi(x^\mu)$ in any given theory $\mathcal{L}$, and for any generator $\epsilon$ and linearized perturbation $\delta \Phi$, the $\boldsymbol{k}_\epsilon(\Phi,\delta\Phi)$ can be found. Then, $\oint_{\partial\Sigma}\boldsymbol{k}_{\epsilon}(\delta\Phi,\Phi)$ gives the $\delta H_{\epsilon}(\Phi)$ as the charge variation inside the hypersurface $\Sigma$. If $\partial\Sigma$ is chosen to be the asymptotics, then $\delta H_{\epsilon}$ would be the charge variation associated with the whole geometry.

The charge variation $\delta H_{\epsilon}$ in Eq. \eqref{delta H xi} may or may not be integrable, conserved, and finite. These conditions are fully discussed in the literature (e.g., see \cite{HS:2015xlp}). Here we report only the $\boldsymbol{k}_\epsilon$ for the Lagrangian densities we studied in this paper, which is the most important tensor for performing the calculations. The details can be found in Ref. \cite{Ghodrati:2016vvf}. Let us consider the following Lagrangian density as the theory under consideration:
\begin{align}\label{Lagrangian scalar}
\mathcal{L}=\frac{1}{16\pi}\Big( & f(R,\phi)\!+\mathrm{a} R_{\mu\nu}R^{\mu\nu}\!+\mathrm{b} R_{\mu\nu\alpha\beta}R^{\mu\nu\alpha\beta}\!-\mathrm{c}_{ab}F_{\mu\nu}^a F^{b\, \mu\nu}\!-\!2\mathrm{d}_{_{IJ}}\nabla^\mu\phi^I\nabla_\mu\phi^J \mp 2\boldsymbol{F}^2\Big).
\end{align}
In this Lagrangian,  $R^\mu_{\,\,\nu\alpha\beta}$, $R_{\mu\nu}$, and $R$ are the Riemann tensor, Ricci tensor, and  Ricci scalar, respectively. $F^a=\mrd A^a$ are some Maxwell fields labeled by index $a$. The $\phi^I$ are some scalar fields labeled by $I$, and $\boldsymbol{F}$ is the cosmological field strength. The coefficients $\mathrm{a}(\phi)$, $\mathrm{b}(\phi)$, $\mathrm{c}_{ab}(\phi)$, and $\mathrm{d}_{_{IJ}}(\phi)$ can be arbitrary functions of $\phi^I$. For clarity, let us give a name for each one of the six parts in the Lagrangian, respectively, as:
\begin{equation}
\mathcal{L}=\mathcal{L}_{f}+\mathcal{L}_\mathrm{a}+\mathcal{L}_\mathrm{b}+\mathcal{L}_\mathrm{c}+ \mathcal{L}_\mathrm{d}+\mathcal{L}_{\boldsymbol{F}}.   
\end{equation}
Using the notation $\boldsymbol{k}_\epsilon=\star k_\epsilon$, then $k_\epsilon$ has a contribution from each one of these parts:
\begin{equation}
k^{\mu\nu}_\epsilon=k^{\mu\nu}_{\epsilon f}+k^{\mu\nu}_{\epsilon \mathrm{a}}+k^{\mu\nu}_{\epsilon \mathrm{b}}+k^{\mu\nu}_{\epsilon\mathrm{c}}+ k^{\mu\nu}_{\epsilon\mathrm{d}}+k^{\mu\nu}_{\epsilon\boldsymbol{F}},
\end{equation}
which can be calculated to be found as

{\small
\begin{flalign}
k_{f\,\epsilon}^{\mu\nu}(\delta\Phi,\Phi)&=\dfrac{1}{16 \pi }\Big[\Big( h^{\mu\alpha}\nabla_\alpha\xi^\nu-\nabla^\mu h^{\nu\alpha}\xi_\alpha-\frac{1}{2}h \nabla^\mu\xi^\nu\Big)f'+2\Big(R^{\mu\alpha}\nabla_\alpha h-\nabla_\alpha R h^{\mu\alpha}-R^\mu_{\,\,\alpha}\nabla_\beta h^{\alpha\beta}\nonumber\\
&\hspace*{1cm}-\Box \nabla^\mu h +\nabla_\alpha\nabla^\mu\nabla_\beta h^{\alpha\beta}-\nabla^\mu(R_{\alpha\beta} h^{\alpha\beta})+\frac{1}{2}\nabla^\mu R\, h \Big)\xi^\nu f''\!\!\nonumber\\
&\hspace*{1cm}+\!2(\nabla^\mu \delta \phi^I \!-\! h^\mu_{\,\,\alpha}\nabla^\alpha\phi^I+\frac{1}{2}h\nabla^\mu\phi^I)\xi^\nu\!\frac{\partial f'}{\partial\phi^I}- \delta \phi^I \nabla^\mu\xi^\nu \frac{\partial f'}{\partial \phi^I} \nonumber\\
&\hspace*{1cm}+\Big(R_{\alpha\beta}h^{\alpha\beta}-\nabla_\alpha\nabla_\beta h^{\alpha\beta}
+\Box h\Big)(\nabla^\mu\xi^\nu f''-2\nabla^\mu R \,\xi^\nu f'''-2\nabla^\mu\phi^I\, \xi^\nu \frac{\partial f''}{\partial \phi^I})\nonumber\\
&\hspace*{1cm}+2\delta \phi^I \nabla^\mu \phi^J\, \xi^\nu \frac{\partial^2 f'}{\partial\phi^I\partial \phi^J}+2\delta \phi^I \nabla^\mu R \,\xi^\nu \frac{\partial f''}{\partial \phi^I}\nonumber\\
&\hspace*{1cm} -\big(f'(\nabla_\alpha h^{\mu\alpha}-\nabla^\mu h)-\nabla_\alpha f'h^{\mu\alpha}+\nabla^\mu f' h\big) \xi^\nu \Big]-[\mu\leftrightarrow\nu], \\
\hspace*{0.2cm}\nonumber\\
k_{\text{a}\,\epsilon}^{\mu\nu}(\delta\Phi,\Phi)&=\dfrac{\text{a}}{16 \pi }\Big[\Big(\nabla^\alpha R_\alpha^{\,\,\mu} h-\nabla_\alpha R h^{\mu\alpha}\!-\!\nabla^\mu(R_{\alpha\beta}h^{\alpha\beta})+\nabla^\mu\nabla_\alpha\nabla_\beta h^{\alpha\beta}-\nabla^\mu \Box h\Big)\xi^\nu \!+\! \Big( 2\nabla_\beta R^\mu_{\,\,\alpha}h^{\beta\nu}\nonumber\\
&\hspace*{1cm}-2 R^{\mu\beta}\nabla_\beta h^\nu_{\,\, \alpha}-2\nabla^\mu R_{\alpha\beta}h^{\nu\beta}- \nabla^\mu (\nabla_\alpha\nabla^\nu h -\nabla_\beta \nabla_\alpha h^{\nu\beta} +\Box h^\nu_{\,\,\alpha}-\nabla^\beta \nabla^\nu h_{\alpha\beta}) \nonumber\\
&\hspace*{1cm} +\nabla^\mu R^\nu_{\,\,\alpha} h +2 R^{\mu\beta} \nabla^\nu h_{\alpha\beta}\Big) \xi^\alpha +\Big(\nabla_\alpha \nabla^\mu h - \nabla_\beta \nabla_\alpha h^{\mu\beta}-\nabla^\beta \nabla^\mu h_{\alpha\beta}+\Box h^{\mu}_{\,\,\alpha}\nonumber\\
&\hspace*{1cm} +2(R_{\alpha\beta}h^{\mu\beta}+R^{\mu\beta}h_{\alpha\beta})-R^{\mu}_{\,\,\alpha} h\Big) \nabla^\alpha \xi^\nu \!+\!\frac{2}{\text{a}}\big(\nabla^\mu R^\nu_{\,\,\alpha}\xi^\alpha+ R^{\nu}_{\,\,\alpha}\nabla^\alpha\xi^\mu - \nabla^\alpha R^\nu_{\,\,\alpha}\xi^\mu\big)\frac{\partial \text{a}}{\partial\phi^I}\delta \phi^I \nonumber \\
&\hspace*{1cm} - \big(2R_{\alpha\beta}\nabla^\alpha h^{\beta \mu}-2\nabla_\alpha R^{\mu}_{\,\,\beta}h^{\alpha\beta}\!-\!R^\mu_{\,\,\alpha}\nabla^\alpha h+\nabla^\alpha R^\mu_{\,\,\alpha}h-R_{\alpha\beta}\nabla^\mu h^{\alpha\beta}\!+\!\nabla^\mu R_{\alpha\beta}h^{\alpha\beta}\big) \xi^\nu\Big]\nonumber \\
&\hspace*{1cm} -[\mu\leftrightarrow\nu],\!\!\!\!\! &&
\end{flalign} 
\begin{flalign}
&k_{\text{b}\,\epsilon}^{\mu\nu}(\delta\Phi,\Phi)=\dfrac{\text{b}}{8 \pi }\Big[\Big(2(R^\mu_{\,\,\alpha\beta\gamma}\!\!-\!R^\mu_{\,\,\beta\alpha\gamma})h^{\nu\gamma}\!\!+\! R^{\mu\,\,\,\nu}_{\,\, \alpha\,\, \beta}h \!-\! R^{\mu\,\,\,\nu}_{\,\,\alpha\,\,\gamma}h_\beta^{\,\,\gamma}\!-\! R^{\mu\,\,\,\nu}_{\,\,\beta\,\,\gamma}h_\alpha^{\,\,\gamma} \!-\! \nabla^\mu \nabla_\alpha h^\nu_{\,\,\beta}\!+\!\nabla^\mu \nabla_\beta h^\nu_{\,\,\alpha}\Big)\!\nabla^\beta\xi^\alpha\nonumber\\
&\hspace*{2.9cm} +\Big(R^{\mu\beta}(\nabla_\alpha h^\nu_{\,\,\beta}-\nabla_\beta h^\nu_{\,\, \alpha})+R^{\mu\,\,\,\nu}_{\,\,\beta\,\,\gamma} \nabla^\gamma h_\alpha^{\,\,\beta}+\frac{1}{2}R^{\mu\nu}_{\,\,\,\,\,\alpha\gamma}(\nabla_\beta h^{\beta\gamma}-\nabla^\gamma h)\nonumber\\
&\hspace*{2.9cm} +2(\nabla_\beta R^\mu_{\,\,\alpha}-\nabla^\mu R_{\alpha\beta})h^{\nu\beta}+\nabla^\mu \nabla_\beta\nabla_\alpha h^{\nu\beta}-\nabla^\mu\Box h^\nu_{\,\,\alpha}+\nabla^\mu R^\nu_{\,\,\alpha}h + \nabla^\mu R^\nu_{\,\,\beta} h_\alpha^{\,\,\beta}\nonumber\\
&\hspace*{2.9cm} -\nabla^\mu (R^\nu _{\,\,\beta\alpha\gamma}h^{\beta\gamma})\Big) 2\xi^\alpha+\frac{2}{\text{b}}\big(\nabla^\alpha R^{\mu\nu}_{\,\,\,\,\,\,\alpha\beta}\,\xi^\beta-R^{\mu\alpha\nu\beta}\nabla_\alpha\xi_\beta\big)\frac{\partial \text{b}}{\partial\phi^I}\delta \phi^I \nonumber \\
& \hspace*{2.9cm} - 2\big(\nabla^\nu R^{\mu}_{\,\,\alpha\nu\beta}h^{\alpha\beta}-R^{\mu}_{\,\,\alpha\nu\beta}\nabla^\nu h^{\alpha\beta}\big) \xi^\nu \Big]-[\mu\leftrightarrow\nu], \\
&k_{\text{c}\,\epsilon}^{\mu\nu}(\delta\Phi,\Phi)=\frac{1}{8 \pi }\Big[\big(\frac{-h}{2} \,\mathrm{c}_{ab}\, F^{a\,\mu\nu}\!+\!2\,\mathrm{c}_{ab}\,F^{a\,\mu\sigma}h_\sigma^{\;\;\nu}-\mathrm{c}_{ab}\,\delta F^{a\,\mu\nu}\!-\!\frac{\partial\,\mathrm{c}_{ab}}{\partial \phi^I}\,F^{a\,\mu\nu}\delta\phi^I\big)({\xi}^\alpha A^b_\alpha+\lambda^b)-\nonumber\\
 & \hspace*{2.9cm}\,\mathrm{c}_{ab}\,F^{a\,\mu\nu}\xi^\alpha \delta A^b_\alpha-2\,\mathrm{c}_{ab}\,F^{a\,\alpha\mu}\xi^\nu \delta A^b_\alpha\Big]-[\mu\leftrightarrow\nu]\,, \\
 &k_{\text{d}\,\epsilon}^{\mu\nu}(\delta\Phi,\Phi)=\frac{1}{4\pi }\Big[\xi^\nu\,\mathrm{d}_{_{IJ}}\,\nabla^\mu\phi^I\,\delta\phi^J\Big]-[\mu\leftrightarrow\nu]\, &&
\end{flalign}
\begin{flalign}
&k^{\mu\nu}_{\boldsymbol{F}\epsilon}=\frac{\pm 1}{8\pi (D-2)!}\Big[\big(\frac{-h^\alpha_{\,\,\alpha}}{2}\boldsymbol{F}^{\mu\nu\rho_3\dots \rho_D}+2 h^{\mu\beta} \boldsymbol{F}_\beta{}^{\nu\rho_3\dots\rho_D}-\delta \boldsymbol{F}^{\mu\nu\rho_3\dots\rho_D}\big) (\xi^\sigma \boldsymbol{A}_{\sigma\rho_3\dots \rho_D}+\boldsymbol{\lambda}_{\rho_3\dots \rho_D})\nonumber\\
&\hspace*{2cm}-\boldsymbol{F}^{\mu\nu \rho_3\dots \rho_D} \xi^\sigma \delta \boldsymbol{A}_{\sigma \rho_3\dots \rho_D}+(D-2)h^{\alpha\beta}\boldsymbol{F}_\alpha{}^{\mu\nu\rho_4\dots\rho_D}(\xi^\sigma \boldsymbol{A}_{\sigma\beta\rho_4\dots\rho_D}+\boldsymbol{\lambda}_{\beta\rho_4\dots\rho_D})\nonumber\\
&\hspace*{2cm}+\frac{2}{D-1} \boldsymbol{F}^{\mu\rho_2\dots\rho_D}\xi^\nu \delta \boldsymbol{A}_{\rho_2\dots\rho_D}\Big]-[\mu\leftrightarrow\nu], && \label{k cosmological}
\end{flalign}
}

\noindent with the notation $h^{\mu\nu}={\delta}g^{\mu\nu}\equiv g^{\mu\alpha}g^{\nu\beta}{\delta}g_{\alpha\beta}$,  ${\delta}\boldsymbol{F}^{\rho_1\dots\rho_D}{\equiv}g^{\rho_1\mu_1}\dots g^{\rho_D\mu_D}{\delta}\boldsymbol{F}_{\mu_1\dots\mu_D}$ and $\delta F^{\mu\nu}\equiv g^{\mu\alpha}g^{\nu\beta}\delta F_{\alpha\beta}$ for the metric, cosmological and Maxwell field strength variations respectively. Besides, the notations $h\equiv h^\mu_{\,\mu}$ and ${\small f'\equiv \frac{\partial f}{\partial R}}$ have been used. We notice that the cosmological gauge field appears explicitly in Eq. \eqref{k cosmological} and its gauge fixing is important for calculation of charges like mass. 

\section{How to check the first law and the Smarr formula if $r_\H$ is not known}\label{app first law Smarr}

Whenever the $r_\H$ is not known in terms of the parameters of the solution $p_i$, one may find checking the Smarr formula and the first law to be difficult, because the entropy is usually an explicit function of $r_\H$. Here, we describe how to check these equations, for black hole solutions whose $r_\H$ is not explicitly known in terms of the free parameters $p_i$ of the solution. The horizon radii are the roots of the equation $\Delta_r\equiv g^{rr}=0$. In order to check the Smarr formula, instead of solving $\Delta_r=0$ to find $r_\H$ as a function of $p_i$, one can solve this equation to find the parameter $m$ as a function of the $\{r_\H,\tilde p_i\}$, which is simpler to be solved. By $\tilde p_i$ we mean all  parameters $p_i$ except the $m$. Then, in the Smarr formula, the parameter $m$ is replaced by its dependency on $\{r_\H,\tilde p_i\}$, and the formula can be checked to hold or not. In order to check the first law, in addition to this procedure, one need to know variations of $r_\H$ with respect to the parameters $p_i$, i.e. the $\delta_{p_i}r_\H$. This can also be found easily by the relation $\delta_{p_i}\Delta_r=0$ (at the horizon), which provides $\delta_{p_i}r_\H=-\frac{\partial \Delta_r}{\partial{p_i}}/\frac{\partial \Delta_r}{\partial r}$ calculated on the horizon (so $r$ will be replaced eventually by $r_\H$).

\end{document}